\documentclass{article}
\usepackage{arxiv}

\usepackage[numbers]{natbib}

\usepackage[british]{babel}
\usepackage[utf8]{inputenc}
\usepackage{textcomp}
\usepackage{gensymb}
\usepackage[acronym]{glossaries}
\usepackage{color}
\usepackage{subcaption}
\usepackage{float}
\usepackage{stfloats}
\usepackage{units}
\usepackage{amsmath,amsthm,mathtools}
\usepackage{amssymb}
\usepackage{multicol}
\usepackage{cuted}
\usepackage{xcolor}

\newacronym[description=Reference scenario]{ref}{Ref}{Reference scenario}
\newacronym[description=Reduced H\textsubscript{2} in EU scenario]{RedH2}{RedH\textsubscript{2}}{Reduced H\textsubscript{2} in EU scenario}
\newacronym[description=Low Russian import scenario]{rus}{LowRus}{Low Russian import scenario}

\newacronym[description=Liquid natural gas]{lng}{LNG}{liquid natural gas}
\newacronym[description=Synthetic natural gas]{sng}{SNG}{synthetic natural gas}

\newacronym[description=Linear Programming]{lp}{LP}{linear programming}
\newacronym[description=Mixed-Integer Programming]{mip}{MIP}{mixed-integer programming}
\newacronym[description=Mixed-Integer Linear Programming]{milp}{MILP}{mixed-integer linear programming}
\newacronym[description=Quadratic Programming]{qp}{QP}{quadratic programming}

\newacronym{eu}{EU}{European Union}
\newacronym{ec}{EC}{European Commission}

\newacronym[description=Power-to-liquid]{ptl}{PtL}{power-to-liquid}
\newacronym[description=Power-to-methane]{ptch4}{PtCH\textsubscript{4}}{power-to-methane}
\newacronym[description=Liquid synthetic natural gas]{lsng}{L-SNG}{liquid synthetic natural gas}

\usepackage{graphicx}
\usepackage{adjustbox}
\usepackage{hyperref}
\usepackage{cleveref}

\usepackage{tabularx}
\usepackage{booktabs}
\usepackage{framed} 
\usepackage{multicol} 
 
\usepackage{nomencl} 
\makenomenclature
 
\usepackage{etoolbox}
\renewcommand\nomgroup[1]{%
  \item[\bfseries
  \ifstrequal{#1}{A}{Sets and indices}{%
  \ifstrequal{#1}{B}{Parameters}{%
  \ifstrequal{#1}{C}{Decision variables}{}}}%
]}

\newcommand{\reals}{{\mbox{\bf R}}}

\newif\ifuniqueAffiliation
\uniqueAffiliationtrue

\ifuniqueAffiliation 
\usepackage{authblk}

\newbox{\orcid}\sbox{\orcid}{\includegraphics[scale=0.06]{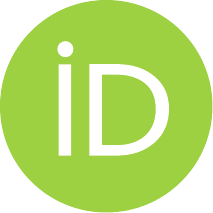}} 
\author[1]{%
	\href{https://orcid.org/0000-0002-1804-4509}{\usebox{\orcid}\hspace{1mm}Felix Frischmuth}%
}
\author[1]{%
	\href{https://orcid.org/0009-0000-6153-2121}{\usebox{\orcid}\hspace{1mm}Mattis Berghoff}%
}
\author[1,2]{%
	\href{https://orcid.org/0000-0003-0857-6760}{\usebox{\orcid}\hspace{1mm}Martin Braun}%
}
\author[1,2]{%
	\href{https://orcid.org/0000-0002-9706-1007}{\usebox{\orcid}\hspace{1mm}Philipp Härtel}%
}
\affil[1]{Fraunhofer Institute for Energy Economics and Energy System Technology IEE, Joseph-Beuys-Str. 8, Kassel, 34117, Hessen, Germany}
\affil[2]{Energy Management and Power System Operation, University of Kassel, 3Wilhelmshöher Allee 73, Kassel, 34121, Hessen, Germany}
\fi

\renewcommand{\shorttitle}{Seasonal hydrogen storage demands under cost and market uptake uncertainties}

\title {Quantifying seasonal hydrogen storage demands under cost and market uptake uncertainties in energy system transformation pathways}   

\begin{document}

\let\WriteBookmarks\relax
\def\floatpagepagefraction{1}
\def\textpagefraction{.001}

\maketitle

\begin{abstract}
Climate neutrality paradigms put electricity systems at the core of a clean energy supply.
At the same time, indirect electrification, with a potential uptake of hydrogen or derived fuel economy, plays a crucial role in decarbonising the energy supply and industrial processes.
Besides energy markets coordinating the transition, climate and energy policy targets require fundamental changes and expansions in the energy transmission, import, distribution, and storage infrastructures.
While existing studies identify relevant demands for hydrogen, critical decisions involve imports versus domestic fuel production and investments in new or repurposing existing pipeline and storage infrastructure.
Linking the pan-European energy system planning model SCOPE SD with the multi-period European gas market model IMAGINE, the case study analysis and its transformation pathway results indicate extensive network development of hydrogen infrastructure, including expansion beyond refurbished methane infrastructure.
However, the ranges of future hydrogen storage costs and market uptake restrictions expose and quantify the uncertainty of its role in Europe's transformation.
The study finds that rapidly planning the construction of hydrogen storage and pipeline infrastructure is crucial to achieving the required capacity by 2050.
\end{abstract}


\section{Introduction}
\label{sec:Introduction}

Clean hydrogen (H\textsubscript{2}) has assumed great importance under the REPowerEU strategy as part of the European Union's new energy security policy \cite{EUROPEANCOMMISSION.2022}.
It plans an ambitious ramp-up of the H\textsubscript{2} economy in Europe over the next few years, both on Europe's consumer and producer side.
Energy sovereignty is receiving increased attention, which implies less import of fossil energy sources and a diverse import mix of renewable-based fuels such as green H\textsubscript{2}.
The European Commission accelerates its progress towards Europe becoming the first climate-neutral continent~\cite{EUROPEANCOMMISSION.2019b}.
To that end, implementing a trans-European energy network is a key element as well as other parts of an infrastructure required for renewable fuels, including new and repurposed H\textsubscript{2} pipeline networks and large-scale electrolyser and storage facilities~\cite{EUROPEANCOMMISSION.2020c}.
\par
The following article focuses on the demands for H\textsubscript{2} storage as a critical element for providing Europe's energy security in the clean transition.
The model-based approach analyses the interactions of an integrated pan-European power and energy system with the trans-European energy infrastructure investments in H\textsubscript{2} storage and pipeline networks.

\subsection{Previous work}
\label{sec:literatureReview}

Recent studies identify significant H\textsubscript{2} demands in deep decarbonisation pathways for Europe, see e.g. \cite{Lux.2022}, and highlight cost benefits from integrated planning of electricity and gas infrastructure and markets, see e.g. \cite{Neumann.2023}. 
While many studies focus on H\textsubscript{2} transport within Europe, e.g. \cite{Lux.2022,v.MikuliczRadecki.2023,Neumann.2023,Backbone.2020}, future needs for large-scale H\textsubscript{2} storage have received less attention in the recent literature.
\par
Studies on the H\textsubscript{2} storage potential exhibit considerable theoretical potentials in salt caverns. 
The work in \cite{Caglayan.2020} indicates a large but unevenly distributed theoretical potential across Europe. 
The analysis in \cite{Lux.2022} determines relevant H\textsubscript{2} storage demands, but does not explicitly include them in the optimisation procedure of the underlying planning problem.  
While a need for seasonal storage is identified, see e.g. \cite{Langfristszenarien.2022}, it remains much lower than today's demands for methane (CH\textsubscript{4}) storage.
The studies in \cite{Neumann.2023}, \cite{Kountouris.2023}, and \cite{Ikaheimo.2023} also point to the importance of H\textsubscript{2} storage for the design
of a future European H\textsubscript{2} infrastructure but focus their analysis on other market and system components.
Furthermore, the models either work with a coarser temporal resolution and less precise regional data or consider a single planning period or year instead of a path-dependent transition.

\subsection{Uncertainty of gas storage demands}
Analysing the increasing coordination of power and gas markets in net-neutral systems requires high temporal granularity, pathway-dependent multi-period analysis, interacting H\textsubscript{2} and CH\textsubscript{4} infrastructure, and multiple sourcing strategies, which have not been the focus of previous models. 
The main contributions of this work include the following:
\begin{itemize}
    \item path-dependent investment decisions for  H\textsubscript{2} and CH\textsubscript{4} infrastructure components as well as system operation decisions of H\textsubscript{2} and CH\textsubscript{4} markets in Europe;
    \item analysis of trade-off between storage and pipeline in Europe's transition to climate neutrality;
    \item coupling of electricity and H\textsubscript{2} markets on a European and global scale with the established energy system planning framework SCOPE SD;
    \item transformation path results for a combination of storage cost uncertainties and market growth variations on European and country level.
\end{itemize}
\par
 The remainder is organised as follows.
 \Cref{sec:methodology} explains the model linking approach and shows a detailed formulation of the IMAGINE framework and the model improvements.
 \Cref{sec:caseStudyDescription} sets up the case study's structure and assumptions for analysing gas sector transformations towards a carbon-neutral Europe.
 \Cref{sec:results} presents and discusses the optimisation model results.
 \Cref{sec:summaryAndConclusion} closes with a summary and draws relevant conclusions.

\section{Methodology}
\label{sec:methodology}

\subsection{Conceptual approach}
\begin{figure}[h]
    \centering
    \includegraphics[width=0.8\columnwidth]{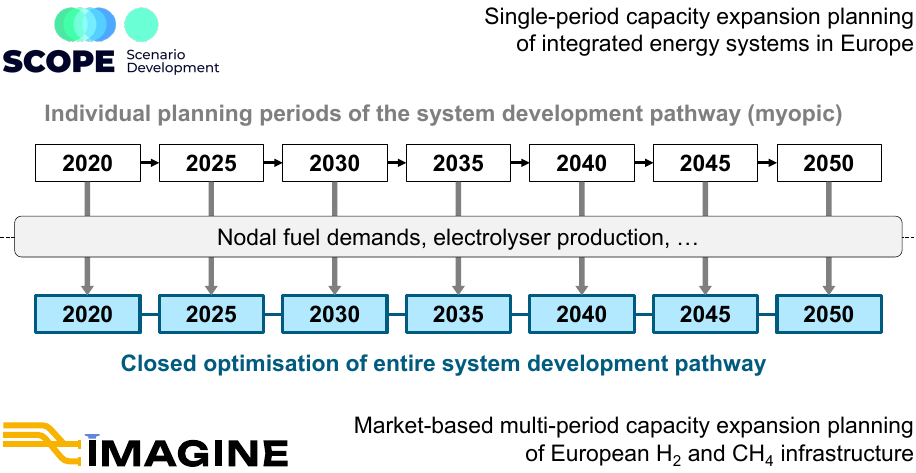}
    \caption{Linking of the single-period integrated energy system model SCOPE SD and the multi-period gas market and infrastructure planning model IMAGINE, own illustration.}
    \label{fig:caseStudyModelLinkage}
\end{figure}

The methodology follows the approach described in \cite{Frischmuth.2022b}, which involves the pan-European cross-sectoral capacity expansion planning framework SCOPE SD (\underline{S}cenario \underline{D}evelopment).
We refer to \cite{Hartel.2020,Hartel.2021,Bottger.2021,Frischmuth.2022,Schmitz.2023} for recent formulations and applications.
The energy system model SCOPE SD is combined with the market-based expansion planning framework IMAGINE (\underline{I}nfrastructure and \underline{ma}rket transformations for \underline{G}as \underline{in} \underline{E}urope). 
Linking those two models involves sectoral H\textsubscript{2} and CH\textsubscript{4} demands, hourly H\textsubscript{2} and CH\textsubscript{4} demand profiles, and H\textsubscript{2} production schedules from domestic electrolysers, all endogenous and consistent results of the pan-European SCOPE SD instances.
\Cref{fig:caseStudyModelLinkage} illustrates the methodology and how both model frameworks are linked in this approach.

Here, we perform one iteration between the two planning models, i.e. SCOPE SD develops a scenario pathway that IMAGINE uses as input.
The sensitivity analysis involves input parameter variations for the IMAGINE model.

\subsection{IMAGINE model}
While SCOPE SD uses only a simplified representation of gas and fuel infrastructure developments, IMAGINE fills that gap by explicitly focusing on production, transport, and storage developments, given the low-carbon transition pathways involving domestic production and renewable import possibilities.
To that end, IMAGINE uses structural and time-series data inputs to minimise all costs incurred for the investment and operation of pipelines, storage facilities, and terminals, as well as the import and domestic production of renewable and low-carbon fuels.
The IMAGINE framework is a bot\-tom-up techno-economic system model that makes deterministic multi-period capacity expansion and system operation decisions for a scenario pathway.
Currently, each country is represented by one node, and the system operation formulations correspond to economic dispatch problems.
Besides national day-ahead H\textsubscript{2} and CH\textsubscript{4} markets that are integrated through cross-border exchange, the model captures different national and global H\textsubscript{2} and CH\textsubscript{4} production options.
By modelling national and pan-European markets, supply decision options include fossil fuels or synthetic renewable energy carriers imported or produced domestically.
The modelling framework further allows the implementation of policy targets and instruments.
\par
The IMAGINE framework is formulated as a \gls{lp} optimisation model in the Python-based Pyomo package~\cite{hart2011pyomo,bynum2021pyomo} and developed as a multi-period decision model to account for pathway dependencies.
In addressing the complexities of the European gas infrastructure transformation, the model adopts a simplified model for gas transport, specifically omitting the intricacies of linepacking, and utilises a coarser temporal modelling resolution, i.e. daily instead of hourly clearing.
The methodological choice is driven by the need to maintain computational tractability when analysing the system-wide dynamics of storage, pipelines, and imports in a multi-period transformation pathway.
Using a coarser temporal and technical resolution in our modelling approach, we focus on the long-term strategic development of the gas infrastructure.
The approach is deemed adequate for capturing the essential interactions and trends at a pan-European level, which is crucial for guiding strategic decisions and policy formulation.
\par
Based on the set definitions in the nomenclature, the model formulation features different parts for the components representing fuel production, pipelines, storage systems, the implemented markets and instruments, and finally, the objective function. 
Note that all decision variables are defined as non-negative reals unless stated otherwise.


\nomenclature[A]{$\Gamma, \gamma$}{Set of planning periods, indexed by $\gamma$}
\nomenclature[A]{$\mathbb{\Gamma}, \gamma$}{Two-dimensional set of investment and operational planning period combinations, indexed by $\gamma = (\gamma^\text{inv},\gamma^\text{op})$}
\nomenclature[A]{$\mathcal{T_\gamma}, t$}{Set of time steps for $\gamma \in \Gamma$, $\mathcal{T}_\gamma = \{ t_1, \dots, t_T \}$, indexed by $t$}
\nomenclature[A]{$\mathcal{I}, i$}{Set of nodes, indexed by $i$}
\nomenclature[A]{$\mathcal{A}^{\text{CH}_4 / \text{H}_2}, a$}{Set of two-dimensional tuples for CH\textsubscript{4}/H\textsubscript{2} pipelines between nodes (directed/undirected arcs), indexed by $a$}
\nomenclature[A]{$\mathcal{S}_i^{\text{CH}_4 / \text{H}_2}, s$}{Set of CH\textsubscript{4}/H\textsubscript{2} storage facilities at node $i$, indexed by $s$}
\nomenclature[A]{$\mathcal{TE}_i^{\text{CH}_4 / \text{H}_2}, te$}{Set of CH\textsubscript{4}/H\textsubscript{2} terminal facilities at node $i$, indexed by $te$}
\nomenclature[A]{$NG_i, g$}{Set of natural gas production facilities at node $i$, indexed by $g$}
\nomenclature[A]{$BH_i$}{Set of blue H\textsubscript{2} production facilities at node $i$, $BH_i \subseteq NG_i$}
\nomenclature[A]{$GH_i, g$}{Set of green H\textsubscript{2} production facilities at node $i$, indexed by $g$}
\nomenclature[A]{$SNG_i, g$}{Set of SNG production facilities at node $i$, $SNG_i \subseteq GH_i$, indexed by $g$}
\nomenclature[A]{$GY_i, g$}{Set of grey H\textsubscript{2} production facilities at node $i$, indexed by $g$}
\nomenclature[A]{$BIO_i, g$}{Set of biomethane production facilities at node $i$, indexed by $g$}
\nomenclature[A]{$DOM_i, g$}{Set of domestic hydrogen production facilities at node $i$, indexed by $g$}

\nomenclature[B]{$RC_g$}{Steam reforming conversion parameter for blue H\textsubscript{2}}
\nomenclature[B]{$V^{{(\cdot)}}_{g,\gamma}$}{Annual natural gas/green hydrogen and \gls{sng} production volume}
\nomenclature[B]{$MC_g$}{\gls{sng} methanation conversion factor}
\nomenclature[B]{$\overline{P}^{\text{CH}_4}_{a,0}$}{Existing CH\textsubscript{4} pipeline capacity}
\nomenclature[B]{$\overline{SV}^{\text{CH}_4}_{s,0}$}{Existing CH\textsubscript{4} storage capacity}
\nomenclature[B]{$\overline{TV}^{\text{CH}_4}_{te,0}$}{Existing LNG terminal capacity}
\nomenclature[B]{$RF_a$}{Effective repurposing factor}
\nomenclature[B]{$AV_{s,t,\gamma}$}{Time-dependent availability factor}
\nomenclature[B]{$MIN_{s}$}{Minimal storage level at storage $s$}
\nomenclature[B]{$\kappa_{s}$}{Storage self-discharge losses between individual time step at storage $s$}
\nomenclature[B]{$\kappa_{s}^\text{WD}$}{Ratio of maximal withdrawal rate to volume of storage $s$}
\nomenclature[B]{$\kappa_{s}^\text{IN}$}{Ratio of maximal injection rate to volume of storage $s$}
\nomenclature[B]{$\lambda_{s}$}{Self-discharge rate of natural gas storages at storage $s$}
\nomenclature[B]{$\eta^{\text{in}}_{s}$}{Injection losses of natural gas storages at storage $s$}
\nomenclature[B]{$\eta^{\text{wd}}_{s}$}{Withdrawal losses of natural gas storages at storage $s$}
\nomenclature[B]{$TLF^{(\cdot)}_{a}$}{Distance-specific losses for CH\textsubscript{4}/H\textsubscript{2} pipelines}
\nomenclature[B]{$D^{(\cdot)}_{i,t,\gamma}$}{CH\textsubscript{4}/H\textsubscript{2} demand}
\nomenclature[B]{$G^{(\cdot)}_{g,t,\gamma}$}{Biomethane or domestic H\textsubscript{2} production}
\nomenclature[B]{$B^{\text{CO}_2}_{\gamma}$}{CO\textsubscript{2} emission budget per period}
\nomenclature[B]{$EF^{(\cdot)}_{g}$}{CH\textsubscript{4}/blue hydrogen/grey hydrogen/LNG emission factor}
\nomenclature[B]{$RF_{te}$}{Reforming conversion parameter for LH\textsubscript{2} terminals at terminal $te$}
\nomenclature[B]{$GR$}{Market growth rate}
\nomenclature[B]{$\Delta_\gamma$}{Years per planning period}
\nomenclature[B]{$\Delta_t$}{Years per time step}
\nomenclature[B]{$PR_{a,\gamma^\text{inv}}$}{Investment restriction for pipelines}
\nomenclature[B]{$SR_{s,\gamma^\text{inv}}$}{Investment restriction for storage}
\nomenclature[B]{$C^{(\cdot)}_{a,\gamma}$}{Lifetime Investment/Repurposing/Decommissioning/ costs parameter for pipelines/storage/terminal}
\nomenclature[B]{$C^\text{op}_{\cdot,\gamma}$}{Operational costs parameter for pipelines/storage/terminals}

\nomenclature[C]{$ng^\text{}_{g,t,\gamma}$}{Natural gas generation}
\nomenclature[C]{$bh^\text{}_{g,t,\gamma}$}{Blue H\textsubscript{2} production}
\nomenclature[C]{$ln^\text{}_{g,t,\gamma}$}{\gls{lng} import}
\nomenclature[C]{$lh^\text{}_{g,t,\gamma}$}{LH\textsubscript{2} import}
\nomenclature[C]{$gh^\text{}_{g,t,\gamma}$}{Green H\textsubscript{2} production}
\nomenclature[C]{$sn^\text{}_{g,t,\gamma}$}{\gls{sng} production}
\nomenclature[C]{$gy_{g,t,\gamma}$}{Grey hydrogen production}
\nomenclature[C]{$p^{{(\cdot)}}_{a=(i,j),t,\gamma} $}{The unidirectional CH\textsubscript{4}/bidirectional H\textsubscript{2} flows from nodes $i \rightarrow j$}
\nomenclature[C]{$\overline{p}^{{(\cdot)}}_{a,\gamma}$}{Available CH\textsubscript{4}/H\textsubscript{2} pipeline capacity}
\nomenclature[C]{$\overline{p}^{{(\cdot)}\uparrow}_{a,\gamma}$}{Newly-built CH\textsubscript{4}/H\textsubscript{2} pipeline capacity}
\nomenclature[C]{$\overline{p}^{{(\cdot)}\downarrow}_{a,\gamma}$}{Decommissioned CH\textsubscript{4}/H\textsubscript{2} pipeline capacity}
\nomenclature[C]{$\overline{p}^{{(\cdot)}\text{,rep}{(\cdot)}}_{a,\gamma}$}{Repurposed pipeline capacity}
\nomenclature[C]{$\overline{sv}^{{(\cdot)}}_{s,\gamma}$}{Available CH\textsubscript{4}/H\textsubscript{2} storage capacity}
\nomenclature[C]{$\overline{sv}^{{(\cdot)}\uparrow}_{s,\gamma}$}{Newly-built CH\textsubscript{4}/H\textsubscript{2} storage capacity}
\nomenclature[C]{$\overline{sv}^{{(\cdot)}\downarrow}_{s,\gamma}$}{Decommissioned CH\textsubscript{4}/H\textsubscript{2} storage capacity}
\nomenclature[C]{$\overline{sv}^{{(\cdot)}\text{,rep}{(\cdot)}}_{s,\gamma}$}{Repurposed storage capacity}
\nomenclature[C]{$\overline{tv}^{{(\cdot)}}_{te,\gamma}$}{Available CH\textsubscript{4}/H\textsubscript{2} terminal capacity}
\nomenclature[C]{$\overline{tv}^{{(\cdot)}\uparrow}_{te,\gamma}$}{Newly-built CH\textsubscript{4}/H\textsubscript{2} terminal capacity}
\nomenclature[C]{$\overline{tv}^{{(\cdot)}\downarrow}_{te,\gamma}$}{Decommissioned CH\textsubscript{4}/H\textsubscript{2} terminal capacity}
\nomenclature[C]{$\overline{tv}^{{(\cdot)}\text{,rep}{(\cdot)}}_{te,\gamma}$}{Repurposed terminal capacity}
\nomenclature[C]{$p^{{(\cdot)}\text{,wd}}_{s,t,\gamma}$}{CH\textsubscript{4}/H\textsubscript{2} withdrawal volume}
\nomenclature[C]{$p^{{(\cdot)}\text{,in}}_{s,t,\gamma}$}{CH\textsubscript{4}/H\textsubscript{2} injection volume}
\nomenclature[C]{$sl_{s,t,\gamma}$}{Storage level}
\nomenclature[C]{$gf_{i,t,\gamma}$}{Natural gas flare}


\printnomenclature

\subsubsection{Preliminary definitions}\label{subsec:preliminary}
Before going into the mathematical formulations for the system components, we define the following sets and mapping functions to allow for concise model formulations:
\begin{subequations}
\begin{gather}
    [k]_j \coloneqq 
\begin{cases}
    \{k,\dots,j\},& \text{if } j\geq k\\
    \varnothing,  & \text{if } j<k
\end{cases} \,,\label{eq:def_list}\\
       \mathbb{\Gamma}^\text{dec} \coloneqq \left \{ (\gamma^\text{inv},\gamma^\text{op}) \in \Gamma \times \Gamma \mid \gamma^\text{op} > \gamma^\text{inv} \wedge \gamma^\text{inv}>1 \right\}\,,\label{eq:def_dec_set}\\
        \mathbb{\Gamma}^\text{rep} \coloneqq \left \{ (\gamma^\text{inv},\gamma^\text{op}) \in \Gamma \times \Gamma \mid \gamma^\text{op} > \gamma^\text{inv} \right\}\,,\label{eq:def_rep_set}\\
        \delta^{\text{CH}_4}(a) \coloneqq \left \{\,(i,j) \in \mathcal{A}^{\text{CH}_4} \mid a = (i,j) \lor a = (j,i) \right\}\,, \label{eq:def_rep_arcs}\\
        \delta^\text{out}(i,\mathcal{A}) \coloneqq \big \{\,a \in \mathcal{A} \mid \exists j \; \text{with} \; a = (i,j) \,\big\}\,, \label{eq:def_delta_out}\\ 
        \delta^\text{in}(i,\mathcal{A}) \coloneqq \big \{\,a \in \mathcal{A} \mid \exists j \; \text{with} \; a = (j,i) \,\big\}\,,\label{eq:def_delta_in}
\end{gather}
\end{subequations}
where \Cref{eq:def_list} defines index lists to map the corresponding multi-period capacity expansion, repurposing, and decommissioning decisions, \Cref{eq:def_dec_set,eq:def_rep_set} define two-dimensional index sets to map the decommissioning and repurposing decisions, \Cref{eq:def_rep_arcs} maps unidirectional CH\textsubscript{4} pipelines to biderectional H\textsubscript{2} arcs, and \Cref{eq:def_delta_out,eq:def_delta_in} define the out- and inflowing arcs for nodal commodity balances.

\subsubsection{Fuel production}
The natural gas generation $ng^\text{}_{g,t,\gamma}$ is limited by the planning period's, e.g. annual, energy production volume $V^{\text{CH}_4}_{g,\gamma}$.
If a facility is viable for blue H\textsubscript{2} production $bh^\text{}_{g,t,\gamma}$, the combined production quantities are limited by that volume, taking the blue H\textsubscript{2} reforming conversion parameter $RC_g \in [0,1]$ into account.
The corresponding constraint writes as follows $\forall g \in {NG}_i$, $\forall i \in \mathcal{I}$, $\forall \gamma \in \Gamma$:
\begin{align}
\sum_{t\in \mathcal{T}_\gamma} \left ( ng^\text{}_{g,t,\gamma} + \sum_{ \substack{\bar{g} \in {BH}_i\\ \bar{g} = g} }  \frac{bh^\text{}_{\bar{g},t,\gamma}}{RC_{\bar{g}}} \right ) \leq V^{\text{CH}_4}_{g,\gamma} \,.
\label{eq:naturalGas}
\end{align}

The production of green H\textsubscript{2} $gh^\text{}_{g,t,\gamma}$ and \gls{sng} $sn^\text{}_{g,t,\gamma}$ are limited by the annual energy production volume $V^{\text{GH}_2}_{g,\gamma}$.
The combined production quantities are limited by that volume, taking the \gls{sng} methanation conversion parameter $MC_g \in [0,1]$ into account.
The corresponding constraint writes as follows $\forall g \in {GH}_i$, $\forall i \in \mathcal{I}$, $\forall \gamma \in \Gamma$:
\begin{align}
\sum_{t\in \mathcal{T}_\gamma} \left ( gh^\text{}_{g,t,\gamma} + \sum_{ \substack{\bar{g} \in {SNG}_i\\ \bar{g} = g} }  \frac{sn^\text{}_{\bar{g},t,\gamma}}{MC_{\bar{g}}} \right ) \leq V^{\text{GH}_2}_{g,\gamma}\,.
\label{eq:greenHydrogen}
\end{align}

\subsubsection{Pipelines}
By explicitly accounting for cross-border exchange flows between bidding zones (nodes), the modelling framework represents market integration and coupling effects.
The pipeline network representation assumes a simplified transport gas flow model and focuses on cross-border exchange, mostly ignoring network constraints arising within bidding zones.
Flows are based on daily capacity values and neglect the gas flow equations, i.e. pressure drop and linepacking in pipeline segments~\cite{Koch.2015}.
\paragraph{Methane (CH\textsubscript{4})}
The available pipeline capacity $\overline{p}^{\text{CH}_4}_{a,\gamma}$ limits the unidirectional CH\textsubscript{4} flows $p^{\text{CH}_4}_{a=(i,j),t,\gamma}$ from nodes $i \rightarrow j$ as follows: 
\begin{subequations}
\begin{gather}
p^{\text{CH}_4}_{a,t,\gamma} \leq  \overline{p}^{\text{CH}_4}_{a,\gamma} \quad \forall a \in \mathcal{A}^{\text{CH}_4},\, \forall t \in \mathcal{T}_\gamma,\, \forall \gamma \in \Gamma \,. \label{eq:capacityConstraintCH4}
\end{gather}
Capacity decisions for the available pipeline capacity are explicitly modelled for every planning period, so that $\forall a \in \mathcal{A}^{\text{CH}_4}$:
\allowdisplaybreaks[3]
\begin{gather}
    \overline{p}^{\text{CH}_4}_{a,1} = \overline{P}^{\text{CH}_4}_{a,0} + \overline{p}^{\text{CH}_4\uparrow}_{a,1} - \overline{p}^{\text{CH}_4\text{,rep}\downarrow}_{a,(1,1)}\,,\label{eq:firstPeriodPipeCH4}\\
    \overline{p}^{\text{CH}_4}_{a,\gamma^\text{op}} = \overline{p}^{\text{CH}_4}_{a,\gamma^\text{op}-1} + \overline{p}^{\text{CH}_4\uparrow}_{a,\gamma^\text{op}} - \sum_{\gamma^\text{inv} \in [1]_{\gamma^\text{op}}} \overline{p}^{\text{CH}_4\text{,rep}\downarrow}_{a,(\gamma^\text{inv},\gamma^\text{op})}
    - \sum_{\gamma^\text{inv} \in [1]_{\gamma^\text{op}-1}} \overline{p}^{\text{CH}_4\downarrow}_{a,(\gamma^\text{inv},\gamma^\text{op})} \qquad\forall \gamma^\text{op} \in \Gamma\setminus{\{1\}}\,,  \label{eq:newPipeCH4}\\
    \overline{p}^{\text{CH}_4\downarrow}_{a,(1,\gamma^\text{op})} \leq \overline{P}^{\text{CH}_4}_{a,0} + \overline{p}^{\text{CH}_4\uparrow}_{a,1} - \sum_{\bar{\gamma}^\text{op} \in [2]_{\gamma^\text{op}-1}} \overline{p}^{\text{CH}_4\downarrow}_{a,(1,\bar{\gamma}^\text{op})}
    - \sum_{\bar{\gamma}^\text{op} = [1]_{\gamma^\text{op}}} \overline{p}^{\text{CH}_4\text{,rep}\downarrow}_{a,(1,\bar{\gamma}^\text{op})} \qquad \forall \gamma^\text{op} \in \Gamma\setminus{\{1\}}\,, \label{eq:decommissioningPipelineFirstPeriodCH4}\\
    \overline{p}^{\text{CH}_4\downarrow}_{a,(\gamma^\text{inv},\gamma^\text{op})} \leq \overline{p}^{\text{CH}_4\uparrow}_{a,\gamma^\text{inv}} - \sum_{\bar{\gamma}^\text{op} \in[\gamma^\text{inv}+1]_{\gamma^\text{op}-1}}   \overline{p}^{\text{CH}_4\downarrow}_{a,(\gamma^\text{inv},\bar{\gamma}^\text{op})}
    - \sum_{\bar{\gamma}^\text{op} = [\gamma^\text{inv}]_{\gamma^\text{op}}} \overline{p}^{\text{CH}_4\text{,rep}\downarrow}_{a,(\gamma^\text{inv},\bar{\gamma}^\text{op})} \qquad \forall (\gamma^\text{inv},\gamma^\text{op}) \in \mathbb{\Gamma}^\text{dec}\,.\label{eq:decommissioningPipelineCH4}
\end{gather}
\end{subequations}
Here, \Cref{eq:firstPeriodPipeCH4} describes the pipeline capacity for CH\textsubscript{4} in the first planning period, which is determined by the existing $\overline{P}^{\text{CH}_4}_{a,0}$, the newly-built $\overline{p}^{\text{CH}_4\uparrow}_{a,1}$, and the repurposed $\overline{p}^{\text{CH}_4\text{,rep}\downarrow}_{a,(1,\gamma^\text{op})}$ (for H\textsubscript{2} transport) capacities in this planning period.
\Cref{eq:newPipeCH4} generalises \Cref{eq:firstPeriodPipeCH4} for the later planning periods, where previous decommissioning and repurposing decisions are taken into account for a given $\gamma^\text{inv}$. 
\Cref{eq:decommissioningPipelineFirstPeriodCH4} limits the decommissioning of existing CH\textsubscript{4} pipeline capacities in the first planning period, and \Cref{eq:decommissioningPipelineCH4} limits the decommissioning of CH\textsubscript{4} pipeline capacities depending on the pre-existing capacity for all other planning periods.
Note that capacities built in the last planning period cannot be decommissioned.
\paragraph{Hydrogen (H\textsubscript{2})}
The model can either build new H\textsubscript{2} or repurpose existing CH\textsubscript{4} pipelines. The corresponding capacity constraints are similar to \Cref{eq:capacityConstraintCH4,eq:firstPeriodPipeCH4,eq:newPipeCH4,eq:decommissioningPipelineFirstPeriodCH4,eq:decommissioningPipelineCH4}.
Unlike CH\textsubscript{4} pipelines, however, H\textsubscript{2} pipelines are assumed to be built in both flow directions.
Hence, cross-border H\textsubscript{2} flows between two nodes $i$ and $j$ can be bidirectional (undirected graph), and are limited by the corresponding installed capacities $\forall a \in \mathcal{A}^{\text{H}_2},\,\forall t \in \mathcal{T},\, \forall \gamma \in \Gamma$:
\begin{subequations}
\begin{gather}
p^{\text{H}_2}_{(i,j),t,\gamma} + p^{\text{H}_2}_{(j,i),t,\gamma} \leq  \overline{p}^{\text{H}_2}_{a,\gamma} \qquad (i,j) =  a\,. 
\end{gather}
Decisions for the available H\textsubscript{2} pipeline capacity $\overline{p}^{\text{H}_2}_{a,\gamma}$ are explicitly modelled for every planning period $\gamma$, so that $\forall a \in \mathcal{A}^{\text{H}_2}$:
\allowdisplaybreaks[3]
\begin{gather}
    \overline{p}^{\text{H}_2}_{a,1} = \overline{P}^{\text{H}_2}_{a,0} + \overline{p}^{\text{H}_2\uparrow}_{a,1} + \overline{p}^{\text{H}_2\text{,rep}\uparrow}_{a,(1,1)} \,,\label{eq:firstPeriodPipeH2}\\
    \overline{p}^{\text{H}_2}_{a,\gamma^\text{op}} = \overline{p}^{\text{H}_2}_{a,\gamma^\text{op}-1} + \overline{p}^{\text{H}_2\uparrow}_{a,\gamma^\text{op}} \nonumber +
    \sum_{\gamma^\text{inv} \in [1]_{\gamma^\text{op}}} \overline{p}^{\text{H}_2\text{,rep}\uparrow}_{a,(\gamma^\text{inv},\gamma^\text{op})}
    - \sum_{\gamma^\text{inv} \in [1]_{\gamma^\text{op}-1}} \overline{p}^{\text{H}_2\downarrow}_{a,(\gamma^\text{inv},\gamma^\text{op})} \quad\forall \gamma^\text{op} \in \Gamma \setminus \{1\}\,, \label{eq:newH2Pipe}\\
    \overline{p}^{\text{H}_2\downarrow}_{a,(1,\gamma^\text{op})} \leq \overline{P}^{\text{H}_2}_{a,0} + \overline{p}^{\text{H}_2\uparrow}_{a,1} + \sum_{\bar{\gamma}^\text{op} = [1]_{\gamma^\text{op}}} \overline{p}^{\text{H}_2\text{,rep}\uparrow}_{a,(1,\bar{\gamma}^\text{op})}
    - \sum_{\bar{\gamma}^\text{op} \in [2]_{\gamma^\text{op}-1}} \overline{p}^{\text{H}_2\downarrow}_{a,(1,\bar{\gamma}^\text{op})} \qquad \forall \gamma^\text{op} \in \Gamma\setminus{\{1\}}\,, \label{eq:decommissioningH2PipelineFirstPeriod}\\
    \overline{p}^{\text{H}_2\downarrow}_{a,(\gamma^\text{inv},\gamma^\text{op})} \leq \overline{p}^{\text{H}_2\uparrow}_{a,\gamma^\text{inv}} + \sum_{\bar{\gamma}^\text{op} = [\gamma^\text{inv}]_{\gamma^\text{op}}} \overline{p}^{\text{H}_2\text{,rep}\uparrow}_{a,(\gamma^\text{inv},\bar{\gamma}^\text{op})}
    - \sum_{\bar{\gamma}^\text{op} \in[\gamma^\text{inv}+1]_{\gamma^\text{op}-1}}  \overline{p}^{\text{H}_2\downarrow}_{a,(\gamma^\text{inv},\bar{\gamma}^\text{op})} \qquad \forall (\gamma^\text{inv},\gamma^\text{op}) \in \mathbb{\Gamma}^\text{dec}\,, \label{eq:decommissioningH2Pipeline}\\
    \overline{p}^{\text{H}_2\text{,rep}\uparrow}_{a,\gamma} = \sum_{\bar{a} \in \delta^{\text{CH}_4}(a)}RF_a\,\overline{p}^{\text{CH}_4\text{,rep}\downarrow}_{\bar{a},\gamma} \quad\forall \gamma \in \mathbb{\Gamma}^\text{rep}\,. \label{eq:repurposedH2Capacity}
\end{gather}
\end{subequations}
Here, \Cref{eq:firstPeriodPipeH2} determines the H\textsubscript{2} transport capacity for the first planning period, which is determined by the newly-built $\overline{p}^{\text{H}_2\uparrow}_{a,\gamma}$, the repurposed $\overline{p}^{\text{H}_2\text{,rep}\uparrow}_{a,(\gamma^\text{inv},\gamma^\text{op})}$, and the decommissioned capacity $\overline{p}^{\text{H}_2\downarrow}_{a,\gamma}$.
Similar to \Cref{eq:newPipeCH4}, \Cref{eq:newH2Pipe} generalises \Cref{eq:firstPeriodPipeH2} for all later planning periods.
Note that incorporating a pre-existing H\textsubscript{2} pipeline capacity $\overline{P}^{\text{H}_2}_{a,0}$ is an option, especially relevant if using the model with a planning horizon starting in planning periods already featuring substantial H\textsubscript{2} infrastructure. 
\Cref{eq:decommissioningH2PipelineFirstPeriod,eq:decommissioningH2Pipeline} limit the decommissioned H\textsubscript{2} pipeline capacities.
\Cref{eq:repurposedH2Capacity} describes the conversion of repurposed CH\textsubscript{4} to H\textsubscript{2} capacities $\overline{p}^{\text{H}_2\text{,rep}\uparrow}_{a,(\gamma^\text{inv},\gamma^\text{op})}$ based on a linear effective repurposing factor $RF_a \in [0,1]$.

\subsubsection{Storage}
The model considers five different types of CH\textsubscript{4} storage components, namely aquifers, rock caverns, salt caverns, depleted fields, and others, of which only salt caverns can be repurposed for H\textsubscript{2} storage.
While capacity expansion and repurposing decisions of gas storage facilities correspond to the pipeline modelling approach, the dispatch constraints differ.
Capacity decisions for the storage (volume) capacity $\overline{sv}^{\text{CH}_4}_{s,\gamma}$ and $\overline{sv}^{\text{H}_2}_{s,\gamma}$ are explicitly modelled for every planning period, see \Cref{eq:firstPeriodStorage,eq:newStorage,eq:decommissioningStorageFirstPeriod,eq:decommissioningStorage,eq:H2Storage,eq:newH2Storage,eq:decommissioningH2Storage,eq:relationRepStroage,eq:repDecommissioningH2Storage}. 
\paragraph{Methane (CH\textsubscript{4})}
Decisions for the available CH\textsubscript{4} storage capacity $\overline{sv}^{\text{CH}_4}_{s,\gamma^\text{op}}$ are explicitly modelled for every planning period and $\forall s \in \mathcal{S}^{\text{CH}_4}$, the constraints are written as follows:
\begin{subequations}
\allowdisplaybreaks[3]
\begin{gather}
    \overline{sv}^{\text{CH}_4}_{s,1} = \overline{SV}^{\text{CH}_4}_{s,0} + \overline{sv}^{\text{CH}_4\uparrow}_{s,1} - \overline{sv}^{\text{CH}_4\text{,rep}\downarrow}_{s,(1,1)}\,, \label{eq:firstPeriodStorage}\\
    \overline{sv}^{\text{CH}_4}_{s,\gamma^\text{op}} = \overline{sv}^{\text{CH}_4}_{s,\gamma^\text{op}-1} + \overline{sv}^{\text{CH}_4\uparrow}_{s,\gamma^\text{op}} - \sum_{\gamma^\text{inv} \in [1]_{\gamma^\text{op}}} \overline{sv}^{\text{CH}_4\text{,rep}\downarrow}_{s,(\gamma^\text{inv},\gamma^\text{op})}
    - \sum_{\gamma^\text{inv} \in [1]_{\gamma^\text{op}-1}} \overline{sv}^{\text{CH}_4\downarrow}_{s,(\gamma^\text{inv},\gamma^\text{op})} \qquad\forall \gamma^\text{op} \in \Gamma\setminus{\{1\}}\,,  \label{eq:newStorage}\\
    \overline{sv}^{\text{CH}_4\downarrow}_{s,(1,\gamma^\text{op})} \leq \overline{SV}^{\text{CH}_4}_{s,0} + \overline{sv}^{\text{CH}_4\uparrow}_{s,1} - \sum_{\bar{\gamma}^\text{op} = [1]_{\gamma^\text{op}}} \overline{sv}^{\text{CH}_4\text{,rep}\downarrow}_{s,(1,\bar{\gamma}^\text{op})}
      - \sum_{\bar{\gamma}^\text{op} \in [2]_{\gamma^\text{op}-1}} \overline{sv}^{\text{CH}_4\downarrow}_{s,(1,\bar{\gamma}^\text{op})}\qquad \forall \gamma^\text{op} \in \Gamma\setminus{\{1\}}\,, \label{eq:decommissioningStorageFirstPeriod}\\
    \overline{sv}^{\text{CH}_4\downarrow}_{s,(\gamma^\text{inv},\gamma^\text{op})} \leq \overline{sv}^{\text{CH}_4\uparrow}_{s,\gamma^\text{inv}} - \sum_{\bar{\gamma}^\text{op} = [\gamma^\text{inv}]_{\gamma^\text{op}}} \overline{sv}^{\text{CH}_4\text{,rep}\downarrow}_{s,(\gamma^\text{inv},\bar{\gamma}^\text{op})}
   - \sum_{\bar{\gamma}^\text{op} \in[\gamma^\text{inv}+1]_{\gamma^\text{op}-1}}   \overline{sv}^{\text{CH}_4\downarrow}_{s,(\gamma^\text{inv},\bar{\gamma}^\text{op})} \qquad \forall (\gamma^\text{inv},\gamma^\text{op}) \in \mathbb{\Gamma}^\text{dec}\,,\label{eq:decommissioningStorage}
\end{gather}
\end{subequations}
The operational constraints characterising the storage systems for CH\textsubscript{4} storage facilities, and they are written as follows $\forall s \in \mathcal{S},\, \forall t \in \mathcal{T},\, \forall \gamma \in \Gamma$:
\begin{subequations}
\begin{gather}
    p^\text{CH\textsubscript{4},wd}_{s,t,\gamma} \leq \textstyle \overline{sv}_{s,\gamma} \, \kappa^{WD}_{s} \, AV_{s,t,\gamma}\,, \label{eq:maxWithdrawal}\\
    p^\text{CH\textsubscript{4},in}_{s,t,\gamma} \leq \textstyle \overline{sv}_{s,\gamma} \, \kappa^{IN}_{s} \, \text{AV}_{s,t,\gamma}\,, \label{eq:maxInjection}\\
    MIN_{s} \, \overline{sv}_{s,\gamma} \, AV_{s,t,\gamma} \leq sl_{s,t,\gamma} \leq \overline{sv}_{s,\gamma} \, AV_{s,t,\gamma}\,, \label{eq:boundsStorageLevel}\\
    sl_{s,t+1,\gamma}= \textstyle sl_{s,t,\gamma} \, \kappa_{s} + \left(p_{s,t,\gamma}^{\text{CH\textsubscript{4},in}} \eta^{\text{in}}_{s} - \frac{p_{s,t,\gamma}^{\text{CH\textsubscript{4},wd}}}{\eta^{\text{wd}}_{s}} \right) \Delta_\text{t} \,,\label{eq:storageContinuity}
\end{gather}
\end{subequations}
where \Cref{eq:maxWithdrawal,eq:maxInjection} limit the maximum storage withdrawal (wd) and injection (in) rates by the planning period-dependent capacities.
\Cref{eq:boundsStorageLevel} limits the storage level including a time-dependent availability parameter $AV_{s,t,\gamma}$ for aggregate maintenance and failure downtimes, and \Cref{eq:storageContinuity} denotes the intra-temporal gas storage continuity for every time step and planning period.
Note that $\kappa_{s} = (1-\lambda_{s})$ accounts for the storage self-discharge losses between individual time steps and that the initial storage level is linked to the last considered time step per planning period $\forall s \in \mathcal{S},\, \forall \gamma \in \Gamma$ by
\begin{equation}
    sl_{s,t_1,\gamma} \leq  \textstyle sl_{s,t_T,\gamma} + \left(p_{s,t_T,\gamma}^{\text{CH\textsubscript{4},in}} \,\eta^{\text{in}}_{s} - \frac{p_{s,t_T,\gamma}^{\text{CH\textsubscript{4},wd}}}{\eta^{\text{wd}}_{s}}\right)\Delta_\text{t}\,.
    \label{eq:storageInitial}
\end{equation}
\paragraph{Hydrogen (H\textsubscript{2})}
Decisions for the available H\textsubscript{2} storage capacity  $\overline{sv}^{\text{H}_2}_{s,\gamma^\text{op}}$ are explicitly modelled for every planning period and written as follows $\forall s \in \mathcal{S}^{\text{H}_2}$.
\begin{subequations}
\allowdisplaybreaks[3]
\begin{gather}
    \overline{sv}^{\text{H}_2}_{s,1} = \overline{SV}^{\text{H}_2}_{s,0} + \overline{sv}^{\text{H}_2\uparrow}_{s,1} + \overline{sv}^{\text{H}_2\text{,rep}\uparrow}_{s,(1,1)} \,,\label{eq:H2Storage}\\
    \overline{sv}^{\text{H}_2}_{s,\gamma^\text{op}} = \overline{sv}^{\text{H}_2}_{s,\gamma^\text{op}-1} + \overline{sv}^{\text{H}_2\uparrow}_{s,\gamma^\text{op}} \nonumber +
    \sum_{\gamma^\text{inv} \in [1]_{\gamma^\text{op}}} \overline{sv}^{\text{H}_2\text{,rep}\uparrow}_{s,(\gamma^\text{inv},\gamma^\text{op})}
    - \sum_{\gamma^\text{inv} \in [1]_{\gamma^\text{op}-1}} \overline{sv}^{\text{H}_2\downarrow}_{s,(\gamma^\text{inv},\gamma^\text{op})} \quad\forall \gamma^\text{op} \in \Gamma \setminus \{1\}\,, \label{eq:newH2Storage}\\
    \overline{sv}^{\text{H}_2\downarrow}_{s,(1,\gamma^\text{op})} \leq \overline{SV}^{\text{H}_2}_{s,0} + \overline{sv}^{\text{H}_2\uparrow}_{s,1}  + \sum_{\bar{\gamma}^\text{op} = [1]_{\gamma^\text{op}}} \overline{sv}^{\text{H}_2\text{,rep}\uparrow}_{s,(1,\bar{\gamma}^\text{op})}
    - \sum_{\bar{\gamma}^\text{op} \in [2]_{\gamma^\text{op}-1}} \overline{sv}^{\text{H}_2\downarrow}_{s,(1,\bar{\gamma}^\text{op})} \qquad \forall \gamma^\text{op} \in \Gamma\setminus{\{1\}}\,, \label{eq:decommissioningH2Storage}\\
    \overline{sv}^{\text{H}_2\downarrow}_{s,(\gamma^\text{inv},\gamma^\text{op})} \leq \overline{sv}^{\text{H}_2\uparrow}_{s,\gamma^\text{inv}} + \sum_{\bar{\gamma}^\text{op} = [\gamma^\text{inv}]_{\gamma^\text{op}}} \overline{sv}^{\text{H}_2\text{,rep}\uparrow}_{s,(\gamma^\text{inv},\bar{\gamma}^\text{op})}
    - \sum_{\bar{\gamma}^\text{op} \in[\gamma^\text{inv}+1]_{\gamma^\text{op}-1}}  \overline{sv}^{\text{H}_2\downarrow}_{s,(\gamma^\text{inv},\bar{\gamma}^\text{op})} \qquad \forall (\gamma^\text{inv},\gamma^\text{op}) \in \mathbb{\Gamma}^\text{dec}\,, \label{eq:repDecommissioningH2Storage}\\
    \overline{sv}^{\text{H}_2\text{,rep}\uparrow}_{s,(\gamma^\text{inv},\gamma^\text{op})} = RF_s\,\overline{sv}^{\text{CH}_4\text{,rep}\downarrow}_{s,(\gamma^\text{inv},\gamma^\text{op})}\quad\forall \gamma\in \mathbb{\Gamma}^\text{rep}\,. \label{eq:relationRepStroage}
\end{gather}
\end{subequations}
The operational constraints characterising the H\textsubscript{2} storage facilities are written as follows $\forall s \in \mathcal{S},\, \forall t \in \mathcal{T},\, \forall \gamma \in \Gamma$:
\begin{subequations}
\begin{gather}
    p^\text{H\textsubscript{2},wd}_{s,t,\gamma} \leq \textstyle \overline{sv}_{s,\gamma} \, \kappa^{WD}_{s} \, AV_{s,t,\gamma}\,, \label{eq:maxWithdrawalH2}\\
    p^\text{H\textsubscript{2},in}_{s,t,\gamma} \leq \textstyle \overline{sv}_{s,\gamma} \, \kappa^{IN}_{s} \, \text{AV}_{s,t,\gamma}\,, \label{eq:maxInjectionH2}\\
    MIN_{s} \, \overline{sv}_{s,\gamma} \, AV_{s,t,\gamma} \leq sl_{s,t,\gamma} \leq \overline{sv}_{s,\gamma} \, AV_{s,t,\gamma}\,, \label{eq:boundsStorageLevelH2}\\
    sl_{s,t+1,\gamma}= \textstyle sl_{s,t,\gamma} \, \kappa_{s} + \left(p_{s,t,\gamma}^{\text{H\textsubscript{2},in}} \eta^{\text{in}}_{s} - \frac{p_{s,t,\gamma}^{\text{H\textsubscript{2},wd}}}{\eta^{\text{wd}}_{s}} \right) \Delta_\text{t} \,,\label{eq:storageContinuityH2}
\end{gather}
\end{subequations}
where \Cref{eq:maxWithdrawalH2,eq:maxInjectionH2} limit the maximum storage withdrawal (wd) and injection (in) rates by the planning period-dependent capacities.
\Cref{eq:boundsStorageLevelH2} limits the storage level including a time-dependent availability parameter $AV_{s,t,\gamma}$ for aggregate maintenance and failure downtimes, and \Cref{eq:storageContinuityH2} denotes the intra-temporal gas storage continuity for every time step and planning period.
Note that $\kappa_{s} = (1-\lambda_{s})$ accounts for the storage self-discharge losses between individual time steps and that the initial storage level is linked to the last considered time step per planning period $\forall s \in \mathcal{S},\, \forall \gamma \in \Gamma$ by
\begin{equation}
    sl_{s,t_1,\gamma} \leq  \textstyle sl_{s,t_T,\gamma} + \left(p_{s,t_T,\gamma}^{\text{H\textsubscript{2},in}} \,\eta^{\text{in}}_{s} - \frac{p_{s,t_T,\gamma}^{\text{H\textsubscript{2},wd}}}{\eta^{\text{wd}}_{s}}\right)\Delta_\text{t}\,.
    \label{eq:storageInitialH2}
\end{equation}

\subsubsection{Markets for CH\textsubscript{4}, H\textsubscript{2}, and CO\textsubscript{2}}  
The market clearing constraints ensure that the demand and supply of CH\textsubscript{4} and H\textsubscript{2} are balanced for each time step.
The market clearing mechanisms are formulated as follows $\forall i \in \mathcal{I},\, \forall t\in \mathcal{T},\, \forall \gamma \in \Gamma$:
\begin{subequations}
\begin{gather}
D^\text{CH\textsubscript{4}}_{i,t,\gamma} = \sum_{g\in NG_i} ng_{g,t,\gamma} + \sum_{te \in \mathcal{TE}^{\text{CH}_4}_i} ln_{te,t,\gamma} + \sum_{g \in SNG_i} sn_{g,t,\gamma} 
- \sum_{g \in GY_i} \frac{gy_{g,t,\gamma}}{\text{RC}_g}  + \sum_{g \in BIO_i} G^\text{BIO}_{g,t,\gamma} \nonumber \\ + \sum_{s\in S^{\text{CH\textsubscript{4}}}_i} \left(p^\text{CH\textsubscript{4},wd}_{s,t,\gamma} - p^\text{CH\textsubscript{4},in}_{s,t,\gamma}\right) + \sum_{a \in \delta^\text{in}(i,\mathcal{A}^{\text{CH}_4})}p^{\text{CH}_4}_{a,t,\gamma} - \sum_{a \in \delta^\text{out}(i,\mathcal{A}^{\text{CH}_4})}p^{\text{CH}_4}_{a,t,\gamma}  TLF^\text{CH\textsubscript{4}}_{a} - gf_{i,t,\gamma} 
\label{eq:marketClearingCH4}\\
D^\text{H\textsubscript{2}}_{i,t,\gamma} = \sum_{g \in BH_i} bh_{g,t,\gamma} + \sum_{g \in \mathcal{TE}^{\text{H}_2}_i} lh_{te,t,\gamma} + \sum_{g \in GH_i} gh_{g,t,\gamma} + \sum_{g \in GY_i} gy_{g,t,\gamma} + \sum_{g \in DOM_i} G^\text{DOM}_{g,t,\gamma} +  \nonumber \\  \sum_{s\in S^{\text{H\textsubscript{2}}}_i  } \left(p^\text{H\textsubscript{2},wd}_{s,t,\gamma} - p^\text{H\textsubscript{2},in}_{s,t,\gamma}\right) + \sum_{a \in \delta^\text{in}(i,\mathcal{A}^{\text{H}_2})}p_{a,t,\gamma} - \sum_{a \in \delta^\text{out}(i,\mathcal{A}^{\text{H}_2})}p_{a,t,\gamma} TLF^\text{H\textsubscript{2}}_{a} \label{eq:marketClearingH2}\,.
\end{gather}
In \Cref{eq:marketClearingCH4}, $D^\text{CH\textsubscript{4}}_{i,t,\gamma}$ is the final (inelastic) CH\textsubscript{4} demand that needs to be covered together with the final (inelastic) biomethane production $G^\text{BIO}_{g,t,\gamma}$.
In \Cref{eq:marketClearingH2}, $D^\text{H\textsubscript{2}}_{i,t,\gamma}$ is the final (inelastic) H\textsubscript{2} demand that needs to be covered together with the final (inelastic) domestic H\textsubscript{2} production $G^\text{DOM}_{g,t,\gamma}$, which can consist of an onshore and offshore part.
In case the desired CO\textsubscript{2} emission instruments include an emission budget and not (only) a price, a separate market is considered $\forall \gamma \in \Gamma$:
\begin{gather}
B^{\text{CO}_{2}}_{\gamma} \geq \sum_{t\in \mathcal{T}} \sum_{i\in \mathcal{I}}\Bigg ( \sum_{g\in NG_i} ng_{g,t,\gamma}  \, EF^{\text{CH}4}_{g}
+ \sum_{g \in BH_i} bh_{g,t,\gamma} \, EF^\text{BH}_{g} \nonumber + \sum_{g \in GH_i} gy_{g,t,\gamma} \, EF^\text{GH}_{g} + \sum_{g \in \mathcal{TE}^\text{CH}_{4}} ln_{g,t,\gamma} \, EF^\text{LNG}_{g} \Bigg) \,,
\label{eq:CO2Budget}
\end{gather}
\end{subequations}
where $B^{\text{CO}_{2}}_{\gamma}$ represents the CO\textsubscript{2} emission budget per planning period.
The system operation decisions must adhere to the planning period-specific budgets.
\par
Note that \Cref{tab:PipelineCosts} contains investment and fixed operation costs for new CH\textsubscript{4} pipelines, new H\textsubscript{2} pipelines and retrofit H\textsubscript{2} pipelines.

\subsection{Model improvement for liquid energy carriers}
To account for the infrastructure needs to import liquid carriers, the original model from \cite{Frischmuth.2022b} has been enhanced to include the possibility of building and decommissioning LNG and LH\textsubscript{2} terminals. 
In addition, existing LNG terminals can be repurposed for LH\textsubscript{2} use. 
The capacity expansion and repurposing decisions correspond to the pipeline modelling approach, see \cite{Frischmuth.2022b}.
\paragraph{Methane (CH\textsubscript{4})}
The decisions for the terminal capacity $\overline{te}^{\text{CH}_4}_{s,\gamma}$ and $\overline{te}^{\text{H}_2}_{s,\gamma}$ are explicitly modelled for every planning period, see \Cref{eq:firstPeriodTerminal,eq:newTerminal,eq:decommissioningTerminalFirstPeriod,eq:decommissioningTerminal,eq:newH2Terminal,eq:decommissioningH2Terminal,eq:repNewH2Terminal,eq:repDecommissioningH2Terminal}.
Decisions for the available CH\textsubscript{4} terminal capacity $\overline{tv}^{\text{CH}_4}_{te,\gamma^\text{op}}$ are explicitly modelled for every planning period, so that $\forall te \in {TE}^{\text{CH}_4}$, $\forall i \in \mathcal{I}$:
\begin{subequations}
\allowdisplaybreaks[4]
\begin{gather}
    \overline{tv}^{\text{CH}_4}_{te,1} = \overline{TV}^{\text{CH}_4}_{te,0} + \overline{tv}^{\text{CH}_4\uparrow}_{te,1} - \overline{tv}^{\text{CH}_4\text{,rep}\downarrow}_{te,1}\,,\label{eq:firstPeriodTerminal}\\
    \overline{tv}^{\text{CH}_4}_{te,\gamma^\text{op}} = \overline{tv}^{\text{CH}_4}_{te,\gamma^\text{op}-1} + \overline{tv}^{\text{CH}_4\uparrow}_{te,\gamma^\text{op}} - \sum_{\gamma^\text{inv} \in [1]_{\gamma^\text{op}}} \overline{tv}^{\text{CH}_4\text{,rep}\downarrow}_{te,(\gamma^\text{inv},\gamma^\text{op})}
    - \sum_{\gamma^\text{inv} \in [1]_{\gamma^\text{op}-1}} \overline{tv}^{\text{CH}_4\downarrow}_{te,(\gamma^\text{inv},\gamma^\text{op})} \qquad\forall \gamma^\text{op} \in \Gamma\setminus{\{1\}}\,,  \label{eq:newTerminal}\\
    \overline{tv}^{\text{CH}_4\downarrow}_{te,(1,\gamma^\text{op})} \leq \overline{TV}^{\text{CH}_4}_{te,0} + \overline{tv}^{\text{CH}_4\uparrow}_{te,1} - \sum_{\bar{\gamma}^\text{op} \in [2]_{\gamma^\text{op}-1}} \overline{tv}^{\text{CH}_4\downarrow}_{te,(1,\bar{\gamma}^\text{op})}
    - \sum_{\bar{\gamma}^\text{op} = [1]_{\gamma^\text{op}}} \overline{tv}^{\text{CH}_4\text{,rep}\downarrow}_{te,(1,\bar{\gamma}^\text{op})} \qquad \forall \gamma^\text{op} \in \Gamma\setminus{\{1\}}\,, \label{eq:decommissioningTerminalFirstPeriod}\\
    \overline{tv}^{\text{CH}_4\downarrow}_{te,(\gamma^\text{inv},\gamma^\text{op})} \leq \overline{tv}^{\text{CH}_4\uparrow}_{te,\gamma^\text{inv}} - \sum_{\bar{\gamma}^\text{op} \in[\gamma^\text{inv}+1]_{\gamma^\text{op}-1}}   \overline{tv}^{\text{CH}_4\downarrow}_{te,(\gamma^\text{inv},\bar{\gamma}^\text{op})}
    - \sum_{\bar{\gamma}^\text{op} = [\gamma^\text{inv}]_{\gamma^\text{op}}} \overline{tv}^{\text{CH}_4\text{,rep}\downarrow}_{te,(\gamma^\text{inv},\bar{\gamma}^\text{op})} \qquad \forall (\gamma^\text{inv},\gamma^\text{op}) \in \mathbb{\Gamma}^\text{dec}\,,\label{eq:decommissioningTerminal}
\end{gather}
\end{subequations}
where the parameter $TV^{\text{CH}_4}_{te,0}$ corresponds to the initial capacity of LNG terminals.
The \gls{lng} imports $ln^\text{}_{g,t,\gamma}$ are limited by the import capacity of $\overline{tv}^{\text{CH}_4}_{te,\gamma^\text{op}}$.
The corresponding constraint writes as follows $\forall te \in \mathcal{TE}^{\text{CH}_4}$, $\forall \gamma \in \Gamma$:
\begin{align}
\sum_{t\in \mathcal{T}_\gamma} ln^\text{}_{te,t,\gamma}  \leq \overline{tv}^{\text{CH}_4}_{te,\gamma^\text{op}} \,.
\end{align}

\paragraph{Hydrogen (H\textsubscript{2})}
Decisions for the available H\textsubscript{2} terminal capacity $\overline{tv}^{\text{H}_2}_{te,\gamma^\text{op}}$ are explicitly modelled for every planning period, so that $\forall te \in \mathcal{TE}^{\text{H}_2}$, $\forall i \in \mathcal{I}$:
\begin{subequations}
    \allowdisplaybreaks[4]
    \begin{gather}
    \overline{tv}^{\text{H}_2}_{te,1} = \overline{TV}^{\text{H}_2}_{te,0} + \overline{tv}^{\text{H}_2\uparrow}_{te,1} + \overline{tv}^{\text{H}_2\text{,rep}\uparrow}_{te,(1,1)} \,,\label{eq:H2Terminal}\\
    \overline{tv}^{\text{H}_2}_{te,\gamma^\text{op}} = \overline{tv}^{\text{H}_2}_{te,\gamma^\text{op}-1} + \overline{tv}^{\text{H}_2\uparrow}_{te,\gamma^\text{op}} \nonumber +
    \sum_{\gamma^\text{inv} \in [1]_{\gamma^\text{op}}} \overline{tv}^{\text{H}_2\text{,rep}\uparrow}_{te,(\gamma^\text{inv},\gamma^\text{op})}
    - \sum_{\gamma^\text{inv} \in [1]_{\gamma^\text{op}-1}} \overline{tv}^{\text{H}_2\downarrow}_{te,(\gamma^\text{inv},\gamma^\text{op})} \quad\forall \gamma^\text{op} \in \Gamma \setminus \{1\}\,, \label{eq:newH2Terminal}\\
    \overline{tv}^{\text{H}_2\downarrow}_{te,(1,\gamma^\text{op})} \leq \overline{TV}^{\text{H}_2}_{te,0} + \overline{tv}^{\text{H}_2\uparrow}_{te,1} - \sum_{\bar{\gamma}^\text{op} \in [2]_{\gamma^\text{op}-1}} \overline{tv}^{\text{H}_2\downarrow}_{te,(1,\bar{\gamma}^\text{op})}
    + \sum_{\bar{\gamma}^\text{op} = [1]_{\gamma^\text{op}}} \overline{tv}^{\text{H}_2\text{,rep}\uparrow}_{te,(1,\bar{\gamma}^\text{op})} \qquad \forall \gamma^\text{op} \in \Gamma\setminus{\{1\}}\,, \label{eq:decommissioningH2Terminal}\\
    \overline{tv}^{\text{H}_2\downarrow}_{te,(\gamma^\text{inv},\gamma^\text{op})} \leq \overline{tv}^{\text{H}_2\uparrow}_{te,\gamma^\text{inv}} - \sum_{\bar{\gamma}^\text{op} \in[\gamma^\text{inv}+1]_{\gamma^\text{op}-1}}  \overline{tv}^{\text{H}_2\downarrow}_{te,(\gamma^\text{inv},\bar{\gamma}^\text{op})}
    + \sum_{\bar{\gamma}^\text{op} = [\gamma^\text{inv}]_{\gamma^\text{op}}} \overline{tv}^{\text{H}_2\text{,rep}\uparrow}_{te,(\gamma^\text{inv},\bar{\gamma}^\text{op})} \qquad \forall (\gamma^\text{inv},\gamma^\text{op}) \in \mathbb{\Gamma}^\text{dec}\,, \label{eq:repDecommissioningH2Terminal}\\
    \overline{tv}^{\text{H}_2\text{,rep}\uparrow}_{te,(\gamma^\text{inv},\gamma^\text{op})} = \sum_{a^\prime \in \delta^{\text{CH}_4}(a)}RF_a\,\overline{tv}^{\text{CH}_4\text{,rep}\downarrow}_{a^\prime,(\gamma^\text{inv},\gamma^\text{op})}\quad\forall \gamma^\text{op} \in \Gamma\,. \label{eq:repNewH2Terminal}
    \end{gather}
    \end{subequations}
\par
The LH\textsubscript{2} imports $lh^\text{}_{g,t,\gamma}$ are limited by the import capacity of a  $\overline{tv}^{\text{H}_2}_{te,\gamma^\text{op}}$.
The corresponding constraint writes as follows $\forall te \in {TE}^{\text{H}_2}$, $\forall \gamma \in \Gamma$:
\begin{align}
\sum_{t\in \mathcal{T}_\gamma} lh^\text{}_{te,t,\gamma} \leq \overline{tv}^{\text{H}_2}_{te,\gamma^\text{op}} \,.
\end{align}

\subsection{Model improvement for market uptake and decommissioning restrictions}
The model aims to optimise cost-efficiency by constructing H\textsubscript{2} and CH\textsubscript{4} infrastructure components as late as possible or repurposing and decommissioning them.  
The improvements address a more realistic system transformation behaviour by considering market growth constraints and avoiding unrealistic overnight build-outs while developing or dismantling gas infrastructure components.

\subsubsection{Market growth rate restrictions}
\label{subsec:marketGrowthRate}
Scaling up projects that construct new or repurpose existing infrastructure components, i.e. pipeline and storage sites, are subject to many potential challenges and restrictions given an initial lack of market maturity.
For instance, adopting and expanding new or reusing existing technologies may encounter high upfront costs with hesitant investment behaviour, compatibility and integration issues, regulatory and compliance hurdles, skill gaps, vendor lock-ins or bottlenecks.
\par
To better understand and address those restrictions in the modelling framework, we introduce an annual market growth rate parameter $GR \in \reals$, such that $GR > 1$, reflecting fast adoption or a slow-growing market environment in the transformation pathway.
The parameter links the capacity expansion decisions of a given planning period with those of the previous planning period, and the corresponding capacity expansion and decommissioning constraints are written as follows $\forall a \in \mathcal{A}$:
\begin{subequations}
\begin{gather}
\overline{p}^{\text{H}_2\uparrow}_{a,\gamma^\text{inv}} \leq \overline{p}^{\text{H}_2\uparrow}_{a,\gamma^\text{inv}-1} GR^{\Delta_\gamma} \quad \forall \gamma^\text{inv} \in \Gamma \setminus \{1\}\,,\\
\sum_{\gamma^\text{op}\in [1]_{\gamma^\text{inv}}} \overline{p}^{\text{H}_2\text{,rep}\uparrow}_{a,(\gamma^\text{inv},\gamma^\text{op})} \leq \sum_{\gamma^\text{op}\in [1]_{\gamma^\text{inv}-1}} GR^{\Delta_\gamma} \overline{p}^{\text{H}_2\text{,rep}\uparrow}_{a,\gamma^\text{inv}-1,\gamma^\text{op})}
\quad \forall \gamma^\text{inv} \in \Gamma\setminus\{1\} \text{ such that } PR_{a,\gamma^\text{inv}} > 0.
\end{gather}
\end{subequations}
The parameter $PR_{a,\gamma^\text{inv}}$ restricts new construction and repurposing of each pipeline by means of a fixed and individual capacity limit for each period. 
As long as $PR_{a,\gamma^\text{inv}} = 0$, new construction and repurposing are completely prohibited. 
The restriction formulated above only begins as soon as these are permitted. 
Storages with the restriction parameter $SR_{a,\gamma^\text{inv}}$ are treated in the same way. 
Note that those parameters are node-specific, which could easily be applied to the entire system. Reasons for this are national differences in workforce and industry.
Also note that the market growth rate only poses a restriction if the growth rate is small enough, e.g. over seven periods, an annual growth rate of $1.02$ implies a doubling of capacity in the last period. 
Assuming a $GR$ value of $1.25$ would allow for more than $2000$ times the initial capacity.
Hence, if the growth rate is large enough, the constraints do not pose a restriction.

\subsubsection{Decommissioning restrictions}
Similarly, the decommissioning decisions of pipeline CH\textsubscript{4} facilities are restricted in the following manner $\forall a \in \mathcal{A}^{\text{CH}_4}$ and $\forall (\gamma^\text{inv},\gamma^\text{op}) \in \mathbb{\Gamma}^\text{dec} \colon \gamma^\text{inv}>\gamma^\text{inv,min}$: 
\begin{equation}
\begin{gathered}
\sum_{\bar{\gamma}^\text{inv} \in[1]_{\gamma^\text{op}-1}}   \overline{p}^{\text{CH}_4\downarrow}_{a,(\bar{\gamma}^\text{inv},\gamma^\text{op})}
\leq 
\sum_{\bar{\gamma}^\text{inv} \in[1]_{\gamma^\text{op}-2}}    GR^{\Delta_\gamma} \overline{p}^{\text{CH}_4\downarrow}_{a,(\bar{\gamma}^\text{inv},\gamma^\text{op}-1)}.
\end{gathered}
\end{equation}
The same approach holds for storage $\forall s \in \mathcal{S}^{\text{CH}_4}$ and $\forall (\gamma^\text{inv},\gamma^\text{op}) \in \mathbb{\Gamma}^\text{dec} \colon \gamma^\text{inv}>\gamma^\text{inv,min}$: 
\begin{equation}
\begin{gathered}
\sum_{\bar{\gamma}^\text{inv} \in[1]_{\gamma^\text{op}-1}}   \overline{sv}^{\text{CH}_4\downarrow}_{s,(\bar{\gamma}^\text{inv},\gamma^\text{op})}
\leq 
\sum_{\bar{\gamma}^\text{inv} \in[1]_{\gamma^\text{op}-2}}    GR^{\Delta_\gamma} \overline{sv}^{\text{CH}_4\downarrow}_{s,(\bar{\gamma}^\text{inv},\gamma^\text{op}-1)}.
\end{gathered}
\end{equation}
The same approach holds for terminals $\forall te \in \mathcal{TE}^{\text{CH}_4}$ and $\forall (\gamma^\text{inv},\gamma^\text{op}) \in \mathbb{\Gamma}^\text{dec} \colon \gamma^\text{inv}>\gamma^\text{inv,min}$: 
\begin{equation}
\begin{gathered}
\sum_{\bar{\gamma}^\text{inv} \in[1]_{\gamma^\text{op}-1}}   \overline{tv}^{\text{CH}_4\downarrow}_{s,(\bar{\gamma}^\text{inv},\gamma^\text{op})}
\leq 
\sum_{\bar{\gamma}^\text{inv} \in[1]_{\gamma^\text{op}-2}}    GR^{\Delta_\gamma} \overline{tv}^{\text{CH}_4\downarrow}_{s,(\bar{\gamma}^\text{inv},\gamma^\text{op}-1)}.
\end{gathered}
\end{equation}

\subsection{Objective and optimisation problem}
The model minimises total costs for investment $f^\text{inv}$ and system operation $f^{\text{op}}$ decisions for all planning periods, covering pipeline, storage and terminal investments as well as operational costs for fuel production and conversion, see \Cref{eq:objectiveValue}.
The investment costs in \Cref{eq:investCosts} consist of the new installation costs $f^\text{new}$ defined in \Cref{eq:newCosts}, the repurposing costs $f^\text{rep}$ defined in \Cref{eq:repurposingCosts}, and the decommissioning costs (including lifetime compensation payments) $f^\text{dec}$ defined in \Cref{eq:decomissiongCosts}. 
The objective function $f$ of the underlying \gls{lp} is:
\allowdisplaybreaks[3]
\begin{subequations}
\begin{gather}
    \min_{(p,sv,tv,ng,bh,ln,lh,gh,gy) \in \mathcal{D}}{f} =  f^{\text{inv}} + f^{\text{op}}\,,
    \label{eq:objectiveValue}  \\
    \mbox{s.t.} \quad \text{Equations (1) to (20)} \nonumber
\end{gather}
where $(p,sv,tv,ng,bh,ln,lh,gh,gy)$ is the tuple of all decision variables and $\mathcal{D}$ its polyhedral feasible set defined by all system-wide and technology-dependent unit constraints formulated above.
The objective function components for investment decisions are defined as
\begin{gather}
    f^\text{inv} = f^\text{new} + f^\text{rep} - f^\text{dec} \,, 
    \label{eq:investCosts} \\
    f^\text{new} = \sum_{\gamma\in \Gamma} \Bigg[ \sum_{a\in \mathcal{A}^{\text{CH}_4}}  C^\text{inv}_{a,\gamma}\overline{p}^{\text{CH}_4\uparrow}_{a,\gamma} + \sum_{a\in \mathcal{A}^{\text{H}_2}}  C^\text{inv}_{a,\gamma}\overline{p}^{\text{H}_2\uparrow}_{a,\gamma}
    + \sum_{s\in \mathcal{S}^{\text{CH}_4}}  C^\text{inv}_{s,\gamma}\overline{sv}^{\text{CH}_4\uparrow}_{s,\gamma} \nonumber \\ + \sum_{s\in \mathcal{S}^{\text{H}_2}}  C^\text{inv}_{s,\gamma}\overline{sv}^{\text{H}_2\uparrow}_{s,\gamma}
    + \sum_{te\in \mathcal{TE}^{\text{CH}_4}}  C^\text{inv}_{te,\gamma}\overline{tv}^{\text{CH}_4\uparrow}_{te,\gamma} + \sum_{te\in \mathcal{TE}^{\text{H}_2}}  C^\text{inv}_{te,\gamma}\overline{tv}^{\text{H}_2\uparrow}_{te,\gamma} \Bigg] \,,
    \label{eq:newCosts}  \\
    f^\text{rep} = \sum_{\gamma \in \mathbb{\Gamma}^\text{rep}} \Bigg[ 
    \sum_{a\in \mathcal{A}^{\text{H}_2}}  C^\text{rep}_{a,\gamma}\overline{p}^{\text{H}_2\uparrow}_{a,\gamma} 
     + \sum_{s\in \mathcal{S}^{\text{H}_2}}  C^\text{rep}_{s,\gamma}\overline{sv}^{\text{H}_2\uparrow}_{s,\gamma}
    + \sum_{te\in \mathcal{TE}^{\text{H}_2}}  C^\text{rep}_{te,\gamma}\overline{tv}^{\text{H}_2\uparrow}_{te,\gamma}
    \Bigg] \,,  
    \label{eq:repurposingCosts} \\
     f^\text{dec} =  \sum_{\gamma \in \mathbb{\Gamma}^\text{dec}} \Bigg[ \sum_{a\in \mathcal{A}^{\text{CH}_4}}  C^\text{dec}_{a,\gamma}\overline{p}^{\text{CH}_4\downarrow}_{a,\gamma} 
    + \sum_{a\in \mathcal{A}^{\text{H}_2}}  C^\text{dec}_{a,\gamma}\overline{p}^{\text{H}_2\downarrow}_{a,\gamma}
    + \sum_{s\in \mathcal{S}^{\text{CH}_4}}  C^\text{dec}_{s,\gamma}\overline{sv}^{\text{CH}_4\downarrow}_{s,\gamma} \nonumber \\ + \sum_{s\in \mathcal{S}^{\text{H}_2}}  C^\text{dec}_{s,\gamma}\overline{sv}^{\text{H}_2\downarrow}_{s,\gamma}
    + \sum_{te\in \mathcal{TE}^{\text{CH}_4}}  C^\text{dec}_{te,\gamma}\overline{tv}^{\text{CH}_4\downarrow}_{te,\gamma} + \sum_{te\in \mathcal{TE}^{\text{H}_2}}  C^\text{dec}_{te,\gamma}\overline{tv}^{\text{H}_2\downarrow}_{te,\gamma}
    \Bigg] \,, 
    \label{eq:decomissiongCosts} 
\end{gather}
where the parameter $C^\text{inv}_{(\cdot),\gamma}$ captures the fixed operation costs and the corresponding specific lifetime investment costs for new capacity investments. Note that these entail lifetime investment and reinvestment costs with a perpetuity approach after the explicitly modelled planning periods.
$C^\text{rep}_{(\cdot),\gamma}$ captures the fixed operation costs and the corresponding specific lifetime investment costs for repurposing capacity investments.
$C^\text{dec}_{(\cdot),\gamma}$ captures the specific decommissioning costs and compensation payments matching the avoided fixed operation costs and specific lifetime investment for decommissioning of capacity.
\Cref{fig:InvesmtentMatrix} illustrates the complex multi-period capacity and investment relations discussed above.
It shows the capacity decisions and costs for three investment streams in a three-period capacity expansion planning instance of the IMAGINE model.
\begin{figure*}[h]
    \centering
    \includegraphics[width=0.9\linewidth]{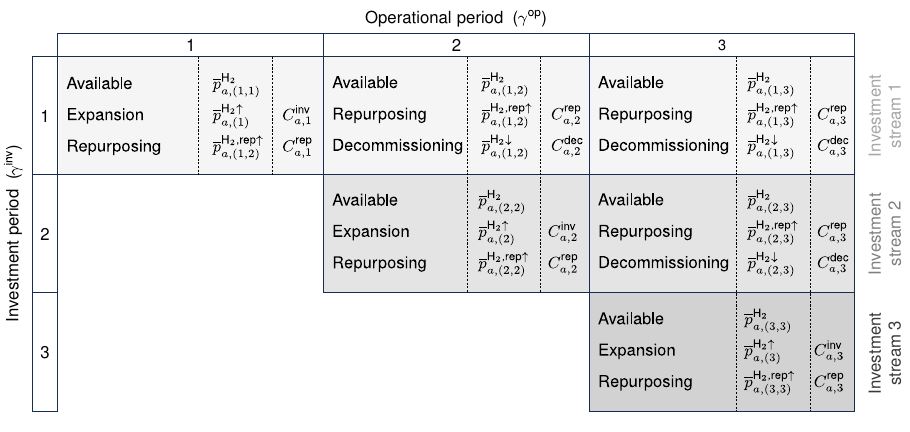}
    \caption{Endogenous capacity decisions and corresponding cost paramters for three investment streams in the multi-period capacity expansion planning model IMAGINE, own illustration.}
    \label{fig:InvesmtentMatrix}
\end{figure*}
The approach offers several advantages when capturing new expansion, decommissioning, and refurbishment decisions in a path-dependent system transition. 
Firstly, it provides a detailed overview of all investment streams, allowing for in-depth analysis of capacity changes, new build-outs, and decommissioning. 
Additionally, it enables consideration of technological improvements in future investment periods, such as enhancements in availability profiles or efficiency of specific technologies. 
Moreover, the method allows for more detailed analysis of repurposing existing or yet-to-be-built capacity.

\par
The objective function components for system operation decisions in every considered time step $t\in T$ are defined as
\begin{gather}
    f^\text{op} = \sum_{\gamma\in \Gamma} \Bigg[ \sum_{t\in \mathcal{T}_\gamma} \bigg (
    \sum_{g \in NG} C^\text{op}_{g,\gamma}ng^\text{}_{g,t,\gamma}
    + \sum_{g \in Bh}  C^\text{op}_{g,\gamma}bh^\text{}_{g,t,\gamma}   
    + \sum_{g \in GH}  C^\text{op}_{g,\gamma}gh^\text{}_{g,t,\gamma}
    + \sum_{g \in SNG} C^\text{op}_{g,\gamma}sn^\text{}_{g,t,\gamma}  
    \nonumber \\ 
    + \sum_{g \in GY} C^\text{op}_{g,\gamma}gy^\text{}_{g,t,\gamma} 
    + \sum_{a\in \mathcal{A}^{\text{CH}_4}} C^\text{op}_{a,\gamma}p^{\text{CH}_4}_{a,t,\gamma} 
    + \sum_{a\in \mathcal{A}^{\text{H}_2}} C^\text{op}_{a,\gamma} \Big( p^{\text{H}_2}_{(i,j),t,\gamma}
    + p^{\text{H}_2}_{(j,i),t,\gamma} \Big)  
    + \sum_{s\in \mathcal{S}^{\text{CH}_4}} C^\text{op}_{s,\gamma} 
    \nonumber \\
    \Big( p^\text{CH\textsubscript{4},wd}_{s,t,\gamma}
    + \, p^\text{CH\textsubscript{4},in}_{s,t,\gamma} \Big) 
    + \sum_{s\in \mathcal{S}^{\text{H}_2}} C^\text{op}_{s,\gamma} \Big( p^\text{H\textsubscript{2},wd}_{s,t,\gamma} + 
    p^\text{H\textsubscript{2},in}_{s,t,\gamma} \Big)  
    + \sum_{te \in \mathcal{TE}^{\text{CH}_4}} C^\text{op}_{te,\gamma}ln^\text{}_{te,t,\gamma} 
    + \sum_{te \in \mathcal{TE}^{\text{H}_2}} C^\text{op}_{te,\gamma}lh^\text{}_{te,t,\gamma}    
    \bigg) \Bigg] \, .
    \label{eq:operationalCosts}
\end{gather}
\end{subequations}
where $C^\text{op}_{(\cdot),\gamma}$ are the operation costs for the corresponding dispatch decision variables for generation and consumption technologies.

\section{Hydrogen storage sensitivity analysis}
\label{sec:caseStudyDescription}
The case study consists of two steps (see \Cref{fig:caseStudyModelLinkage}).
First, SCOPE SD generates medium- and long-term scenarios for the future net-zero European energy system across seven expansion planning periods from 2020 to 2050 in five-year intervals, i.e. $\Gamma = \{1,...,7\}$.
Second, the SCOPE SD results provide input for IMAGINE to examine ten scenario variants that represent the uncertainty of H\textsubscript{2} storage costs and market uptake restrictions in the transformation pathways.

\subsection{Structural and time series input data}
\label{subsec:structuralTimeSereisInputData}
Detailed information on input data and assumptions for the SCOPE SD instances can be found in recent publications, see \cite{Hartel.2020,Hartel.2021,Bottger.2021,Frischmuth.2022}.
The case study setup uses the historical meteorological year 2012 (including its ``Kalte Dunkelflaute'' period) and assumes that Europe is climate-neutral by 2050.
The scenario is based on \cite{Gerhardt.2023} and visualised at Fraunhofer IEE's \textit{Transformationsatlas der Energiewende}~\cite{FraunhoferIEE.2023} (Ariadne Base Scenario).
\par
Running the IMAGINE modelling and optimisation framework requires additional structural and time series input data.
Information on already existing pipelines in 2020 is taken from \cite{ENTSOG.2022} and data on already existing storage capacities in 2020 is taken from \cite{GIE.2021} and for LNG terminals from \cite{GIE.2022}.
Investment and operation costs for future pipelines are based on \cite{Backbone.2020}, for terminals on \cite{FlorianSchreinerMatiaRiemerJakobWachsmuth.} and for storage on \cite{BDI.2022} and \cite{DEA.2020}, see \Cref{tab:PipelineCosts} in the appendix.
As available cost information for decommissioning infrastructure is very limited in the literature, the case study approximates these at 10\,\% of the corresponding investment costs~\cite{Kaiser.2017}. 
The future underground H\textsubscript{2} storage potential including depleted hydrocarbon reservoirs and salt caverns is gathered from~\cite{Caglayan.2020}, which indicates a large but unevenly distributed technical potential of salt caverns across Europe.
\par
Price and quantity assumptions for non-European export countries for renewable energy carriers are based on Fraunhofer IEE's \textit{Power-to-X atlas}~\cite{FraunhoferIEE.2021}. 
Historical data for natural gas generation and gas reserves is taken from \cite{eurostat.2022,BMWK.2021}. 
\Gls{lng} import costs are estimated using \cite{Steuer.2019} and natural gas production costs are determined using \cite{DIWBerlin.2019}.
\par
Time series data of production and consumption as well as demand for H\textsubscript{2} and CH\textsubscript{4} are derived from the upstream SCOPE SD results. To that end, hourly results from SCOPE SD are aggregated to daily input data for the IMAGINE framework.
The carbon price path is interpolated between a price of 50\,EUR/tCO\textsubscript{2} in 2020 \cite{CO2.2021} and 400\,EUR/tCO\textsubscript{2} by 2050, which is an endogenous result of the SCOPE SD model.
\par
Note that all investment decisions are based on an interest rate of 6\,\% and a depreciation of 50\,years for pipelines and 50\,years for storage. 
A social discount rate of 3\,\% is assumed to consider the time value of money when making multi-period investment and decommissioning decisions in the IMAGINE framework.

\subsection{Scenario setup}
Experience with the construction and use of H\textsubscript{2} storage facilities is still limited.
Due to the inherent individuality of each project and the ongoing technology development, it is difficult to forecast the exact costs and market uptake rates of deploying solutions involving this technology.
To better understand the relationship between H\textsubscript{2} storage costs and market uptake rates for potential gas market and infrastructure transformation pathways, we perform a sensitivity analysis based on those two parameters.

\paragraph{Storage cost sensitivity}
Given the current uncertainty associated with H\textsubscript{2} storage investment costs, we explore a wide range of capital expenditure (CAPEX) projections based on very high and low estimates from the recent literature.
More specifically, we assume the following two storage cost scenarios for newly-built H\textsubscript{2} storage facilities:
\begin{itemize}
    \item \textit{Low}: 269,393\,EUR/GWh\textsubscript{th} (BDI)~\cite{BDI.2022},
    \item \textit{High}: 1,500,000\,EUR/GWh\textsubscript{th} (DEA)~\cite{DEA.2020},
\end{itemize}
and add eight intermediate cost steps in between for a storage cost range with a total of 10 variations (see \Cref{tab:PipelineCosts} in the Appendix for detailed information).
Note that the \textit{Medium} storage cost scenario assumes 808,179\,EUR/GWh\textsubscript{th}.
The cost variations apply to CAPEX, fixed operational expenditures (OPEX), and decommissioning costs.
The variable OPEX are not part of the variation as they are not expected to change proportionally with the other economic parameters.

\paragraph{Market growth rate sensitivity}
In a second part, we conduct a sensitivity analysis varying the market growth rate parameters, see \Cref{subsec:marketGrowthRate}.
We analyse the following four different growth rate restrictions: 
\begin{itemize}
    \item $GR=1.05$,
    \item $GR=1.25$,
    \item $GR=1.50$,
    \item $GR=\infty$ (no restriction).
\end{itemize}
While the storage cost sensitivity analysis assumes a constant $GR=1.25$ for the storage cost range, the market growth rate sensitivity analysis also takes storage cost variations into account, i.e. \textit{Low}, \textit{Medium}, and \textit{High}.
In total, both sensitivity analyses result in 19 individual IMAGINE model runs.

\paragraph{Simulation setup}
As mentioned, the IMAGINE framework is formulated as a \gls{lp} optimisation model, and the resulting large-scale \gls{lp} instances are solved with the Barrier (interior point) algorithm of \verb|Gurobi Optimizer| \verb|Version| \verb|11.0|~\cite{gurobi} on a medium-range HPCC node at Fraunhofer~IEE (Intel XEON E5-2698v3 16\,Cores@2.30\,GHz, 256\,GB).
The matrix size is characterised as follows: 9.2\,million variables, 9.2\,million constraints and 38\,million non-zero entries. 
Solving times of the problem instances vary between 45\,min and one hour.  

\section{Results and discussion}
\label{sec:results}
In this section, we analyse the scenario results based on our model.
Our aim is to explore how the uncertainty of costs for H\textsubscript{2} storage and market uptake restrictions impacts the transition in Europe and the long-term H\textsubscript{2} economy.
We present the results in two parts: \Cref{subsec:transformedSystem} shows the overall infrastructure deployments and infrastructure operation in the transformed (finale state) system, and \Cref{subsec:transformationPathways,subsec:HydrogenStroageCostUncertainty,subsec:MarketGrowthUncertainty} illustrates the results of the parameter variations and the system transformation pathways. 

\subsection{Transformed system}
\label{subsec:transformedSystem}

\begin{figure*}[h]
    \centering
    \includegraphics[width=0.98\linewidth]{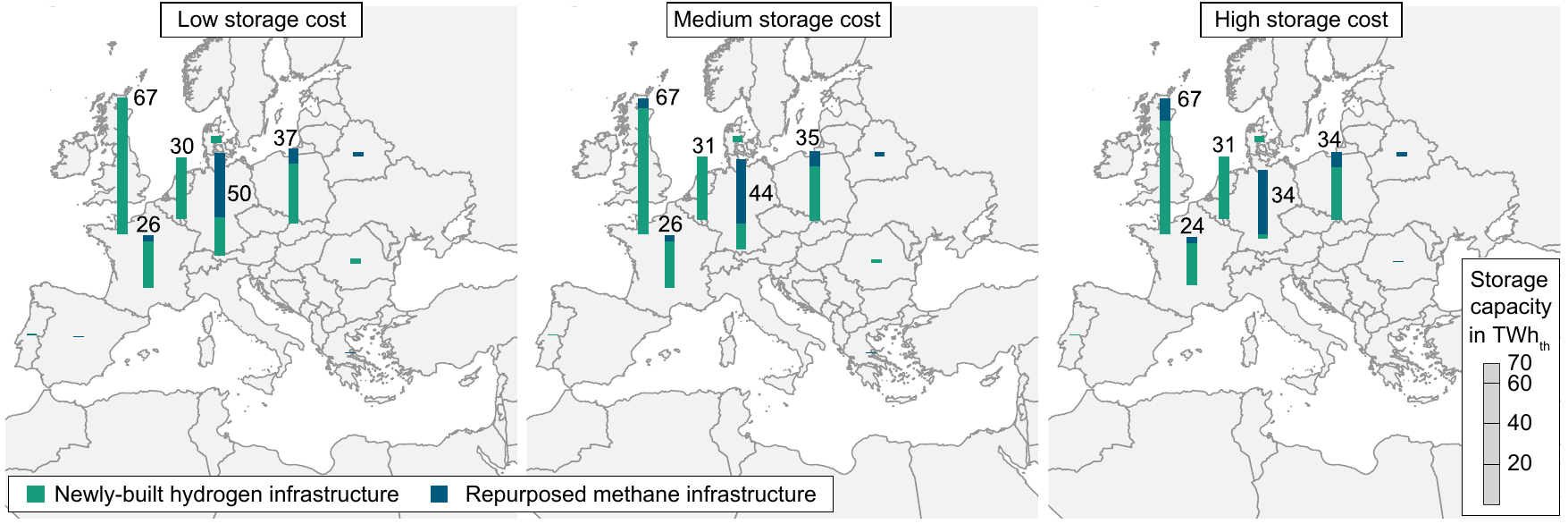}
    \includegraphics[width=0.98\linewidth]{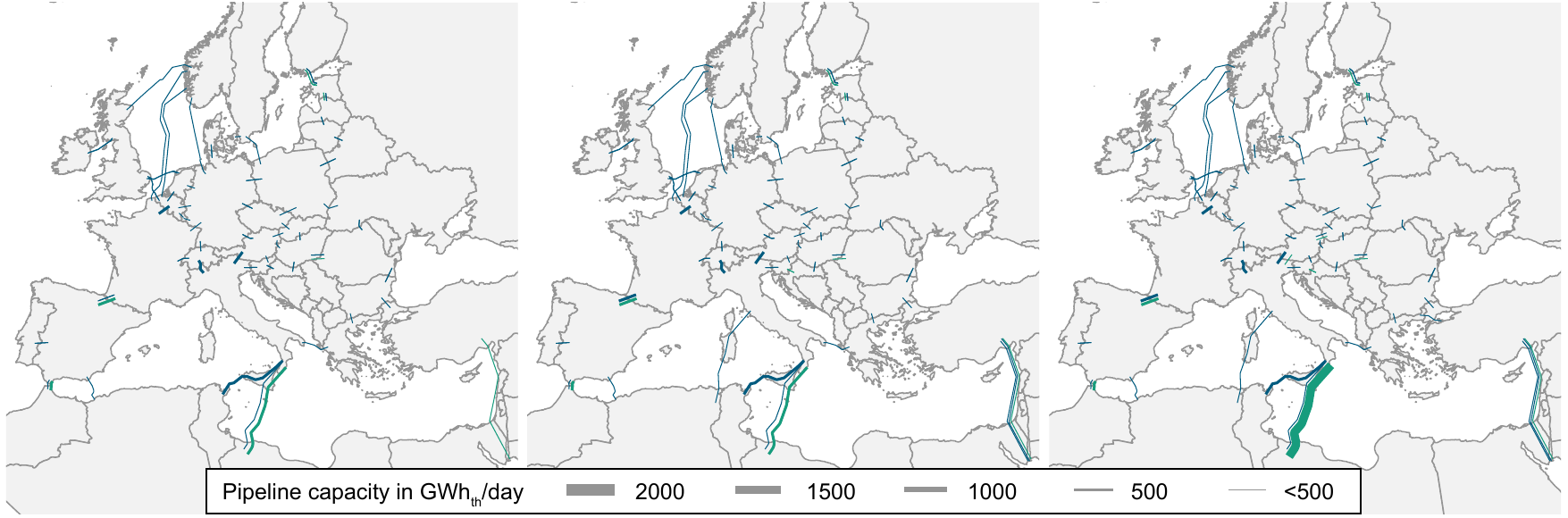}
    \caption{Installed capacities of H\textsubscript{2} storage (top) and pipelines (bottom) across Europe for the low, medium, and high storage cost scenarios in the transformed system by 2050 ($GR=1.25$), own illustration based on optimisation results.}
    \label{fig:europe_map_compasion_2050}
\end{figure*}

\paragraph{Infrastructure demands in the transformed system}
\Cref{fig:europe_map_compasion_2050} illustrates the installed capacities of H\textsubscript{2} storage and pipelines across Europe.
Different H\textsubscript{2} storage costs significantly affect the gas infrastructure development until 2050.
Europe's H\textsubscript{2} storage development concentrates on the salt cavern potential in central Europe, particularly in the market areas of Germany, Poland, France, Netherlands, and Great Britain. 
In the low-cost scenario, we observe substantial installations of new H\textsubscript{2} storage.
Moreover, the results show that, except for the high storage cost scenario, the potential of repurposing CH\textsubscript{4} storage is largely exploited.
At the same time, there are only limited installations of new H\textsubscript{2} storage systems.
\par
Great Britain's continuing natural gas consumption creates challenges in repurposing CH\textsubscript{4} storage as the market growth rate imposes restrictions.
The initial repurposing decisions would have to start at an earlier stage, which is impossible due to the continuing CH\textsubscript{4} consumption.
With increasing storage costs, the model compensates for repurposed natural gas storage with pipelines.
Hence, Great Britain is an outlier in the repurposing of storage facilities and seems to be build out exactly the same amount of storage in all scenarios.
\par
In contrast, Germany experiences the most significant decrease in storage build-out as storage costs increase. 
More than the half of the less installed storage across Europe in 2050 between the low- and high-cost scenario comes from Germany. 
This highlights Germany's role as a central transit country in Europe.
Additional pipeline investments may compensate for the lack of storage infrastructure in Germany. However, this may not be the case for other countries that are less connected in the European H\textsubscript{2} network due to their geographical location.
Considering no network restrictions within a node like Germany also impacts the results for storage needs. 
\par
As expected, pipeline developments have opposite effects.
In the low-cost scenario, the largest capacity consists of repurposed pipeline capacity. 
In the high-cost scenario, new H\textsubscript{2} pipeline capacity is installed to connect the Middle East and North Africa (MENA) region with Central Europe.
This is particularly evident for the pipeline links from Libya to Italy.
The H\textsubscript{2} import capacity grows by 1\,TWh\textsubscript{th}/day, while the pipeline capacity from the other European countries to Italy grows by 0.5\,TWh\textsubscript{th}/day.
For example, this connection can compensate for reduced storage build-outs in Central Europe for the high-cost scenario. 

\begin{figure*}[h]
    \includegraphics[width=.48\linewidth]{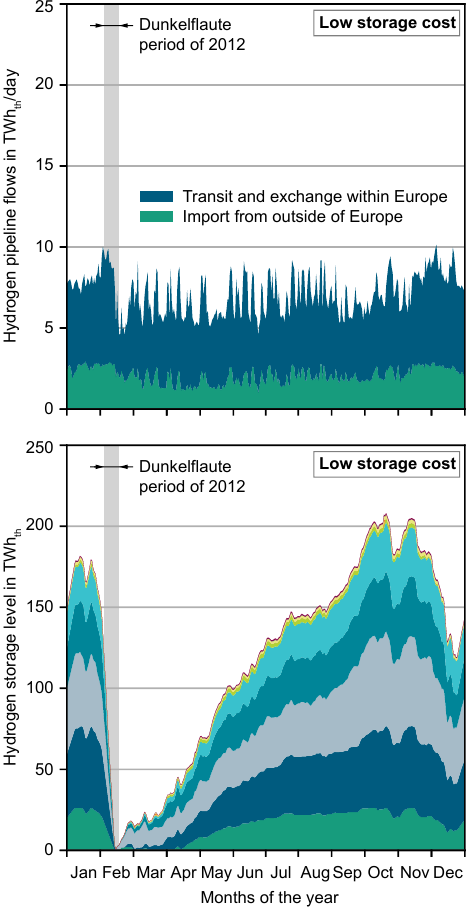}\hfill
    \includegraphics[width=.48\linewidth]{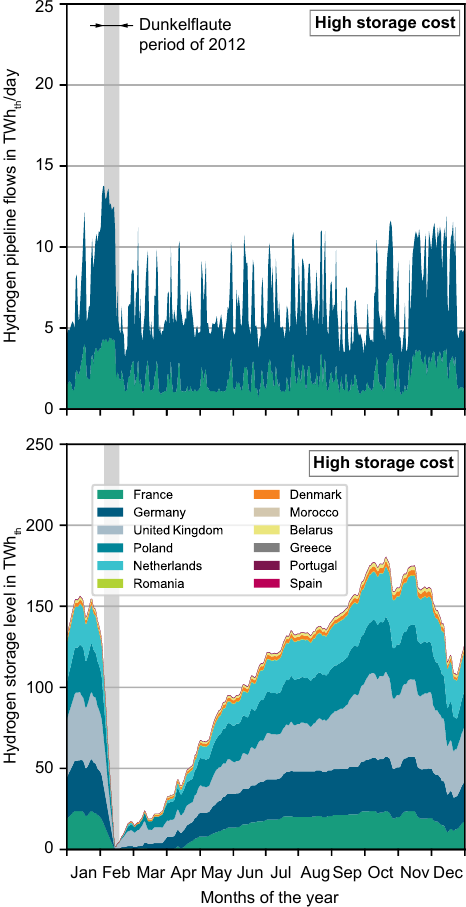}
    \caption{H\textsubscript{2} pipeline flows (top) and H\textsubscript{2} storage (salt cavern) trajectory (bottom) for the low (left) and high (right) storage cost scenarios assuming a market growth rate of $GR = 1.25$, own illustration based on optimisation results.}
    \label{fig:hydrogen_stroage_pipeline_trajh}
\end{figure*}

\paragraph{Infrastructure operation in the transformed system}
A comparison of the daily dispatch results for the entire year of 2050 in the low- and high storage cost scenarios shows that seasonal demand variations are met by either significant storage expansion or more extensive pipeline infrastructure within Europe and from the exporting MENA region to Europe, see \Cref{fig:hydrogen_stroage_pipeline_trajh}).

Note that all storage systems' initial and final volume levels are linked by an inequality constraint, which requires the final volume level to be greater than or equal to the initial volume level. 
Therefore, the observed storage trajectories, including the initial and final volume levels, are endogenous decisions optimised by the model. 

As mentioned in the \Cref{subsec:structuralTimeSereisInputData}, 2012 was chosen as the meteorological input year and the effect of the ``Kalte Dunkelflaute'' is highlighted in \Cref{fig:hydrogen_stroage_pipeline_trajh}.
During the two weeks, it is evident that the energy system suffers from a lack of renewable electricity generation and heat supply coinciding with high energy consumption during cold winter days. 
The European energy system and its security of supply depend on the energy carrier H\textsubscript{2} during that period.  
On the one hand, H\textsubscript{2} storage investments depend on the assumed costs.
On the other hand, working gas injection and withdrawal ratios play a crucial role in the expansion decisions. 
These are also uncertain parameters due to limited knowledge of existing and future storage configurations.
\Cref{tab:storage_expansion} presents the difference between installed capacity, the so-called working gas volume, and the actually used storage volume.
Note that the maximum used volume must be lower than the installed capacity.
The data of existing gas storage facilities are heterogeneous in their technical specification.
While depleted fields or aquifers feature higher working gas to injection or withdrawal design ratios of over 70\,GWh\textsubscript{th}/GW\textsubscript{th}, salt caverns feature around 30 to 50\,GWh\textsubscript{th}/GW\textsubscript{th} (weighted mean), see \cite{GIE.2021}. 
Different assumptions for those parameters significantly influence the installed capacity results.
The results of the mean design ratio for the combination of new build and repurposed storage are given in \Cref{tab:storage_expansion}. 
\begin{table}[h]
    \centering
    \caption{Hydrogen (H\textsubscript{2}) storage and pipeline capacity relation of installed capacity and used capacity in 2050 ($\gamma^\text{op}=7$).}
    \begin{tabular}{l r r}
        \toprule
          Storage cost scenario (2050) & Low & High \\
          \midrule
          Share of repurposed storage in \% & 20.2 & 28.8 \\  
         \midrule
          Installed capacity $\sum_{s\in \mathcal{S}^{\text{H}_2}}  \overline{sv}^{\text{H}_2}_{s,7}$ in TWh\textsubscript{H\textsubscript{2}} & 220.6 & 195.2 \\
          $\max_{t\in \mathcal{T}_{7}}\big(\sum_{s\in \mathcal{S}^{\text{H}_2}}  sl_{s}(t)\big)$ in TWh\textsubscript{H\textsubscript{2}}  & 212.0 & 180.6 \\
          \midrule
          Share of repurposed pipelines in \% & 77.5 & 82.9 \\ 
          \midrule
          Installed capacity $\sum_{a\in \mathcal{A}^{\text{H}_2}}  \overline{p}^{\text{H}_2}_{a,7}$ in TWh\textsubscript{H\textsubscript{2}}/d & 12.5 &  15.4 \\
          $\max_{t\in \mathcal{T}_{7}}\big(\sum_{a\in \mathcal{A}^{\text{H}_2}} {p_a}^{\text{H}_2}\big)$ in TWh\textsubscript{H\textsubscript{2}}/d & 10.1 & 13.8 \\
        \bottomrule
    \end{tabular}
    \label{tab:storage_expansion}
\end{table}

\subsection{Transformation pathways}
\label{subsec:transformationPathways}

\begin{figure*}
    \centering
    \includegraphics[width=\linewidth]{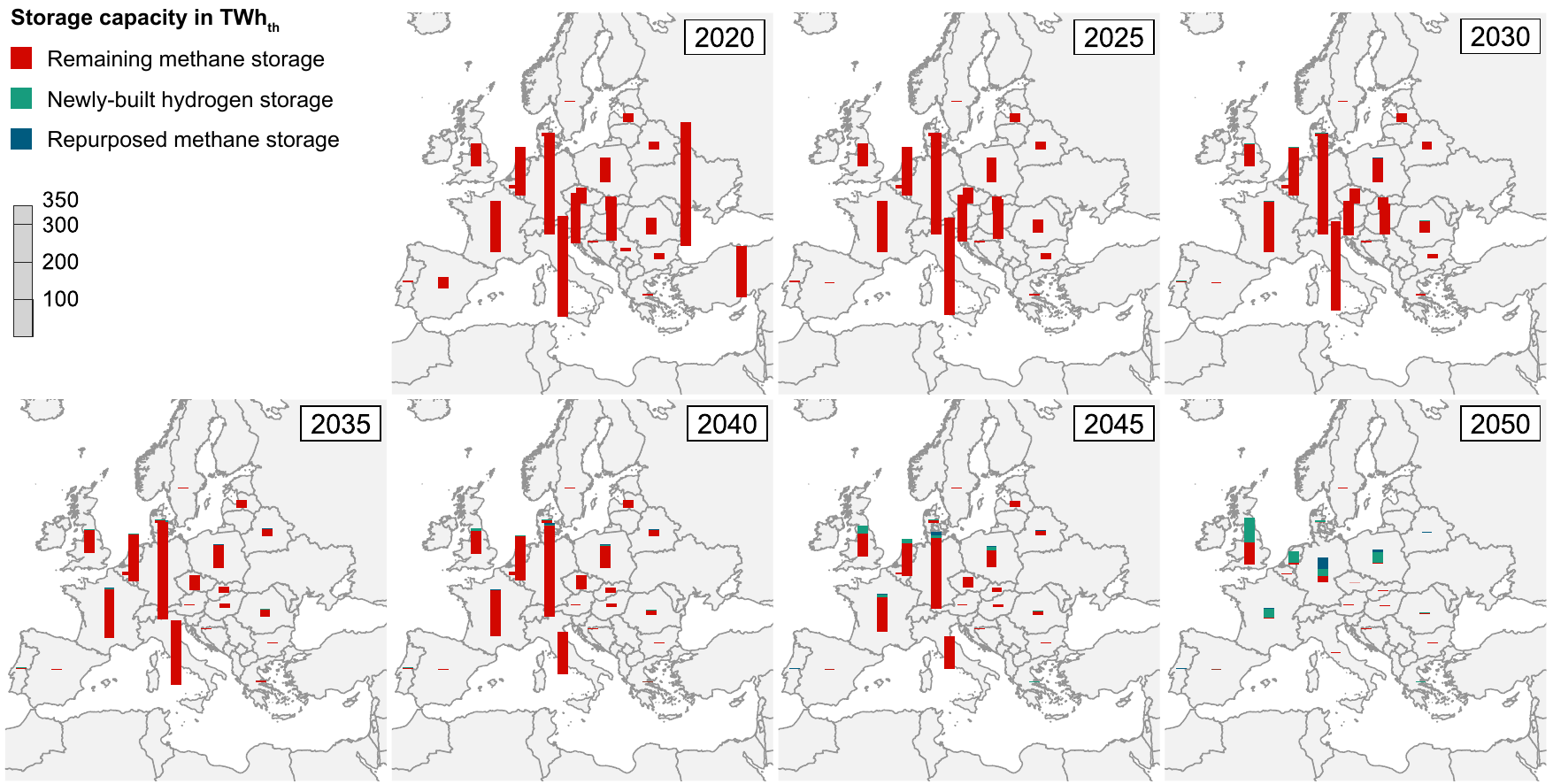}
    \caption{Hydrogen (H\textsubscript{2}) storage expansion pathway for all considered market areas at the low-cost scenario and a market growth rate of $GR = 1.25$, own illustration based on optimisation results.}
    \label{fig:hydrogen_stroage_path}
\end{figure*}

All scenario variants include path-dependent investment and decommissioning decisions.
\Cref{fig:hydrogen_stroage_path} illustrates the path-dependent development of H\textsubscript{2} storage in Europe for the low storage cost scenario with a market growth rate of $1.25$. 
A key result is the late (2045-2050) build-out of storage facilities in Europe, which results from the seasonal demand patterns in the long-term system states with achieved carbon neutrality.
In the medium-term planning periods, i.e. 2030-2045, the less seasonal H\textsubscript{2} demand is met by H\textsubscript{2} exchange via pipelines.
Note that, due to the use of renewable electricity and more efficient technologies, e.g. heat pumps, the demands for H\textsubscript{2} and, ultimately, H\textsubscript{2} storage are smaller compared to today’s CH\textsubscript{4} storage needs.
\par
Great Britain stands out with its continued higher storage demand for CH\textsubscript{4}. 
Possible reasons include the island location, decommissioning constraints enforced by the market growth rate, and a distinct CH\textsubscript{4} consumption pattern including a considerable share of gas-based thermal power plants.
Note that some of arguments also apply to Germany which has a comparatively smaller but still visible sustained demand for CH\textsubscript{4} storage.

\subsection{Impacts of hydrogen storage cost}
\label{subsec:HydrogenStroageCostUncertainty}
\begin{figure}[h]
    \centering
    \includegraphics[width=0.48\columnwidth]{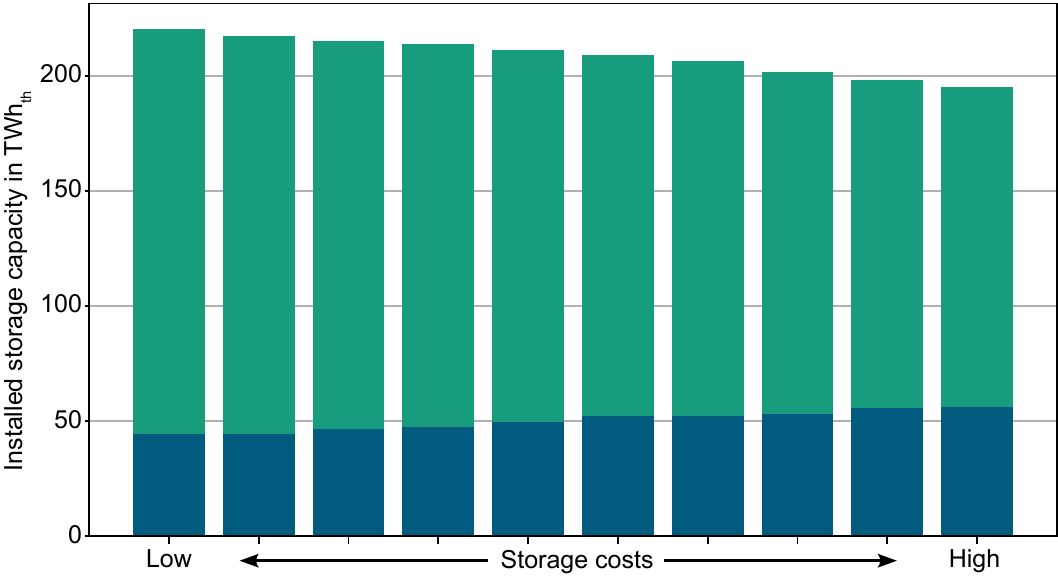}\hfill
    \includegraphics[width=0.48\columnwidth]{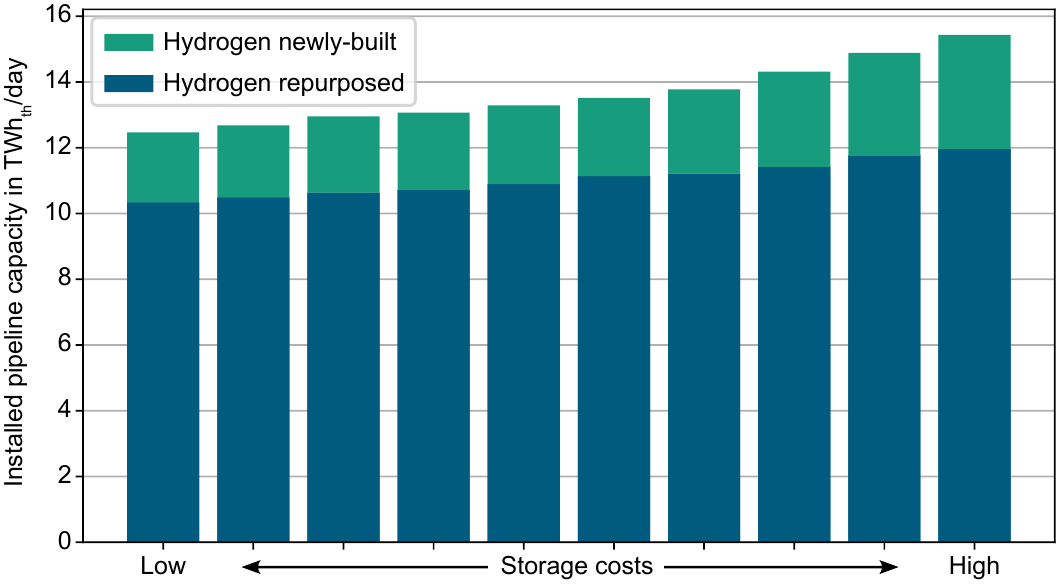}
    \caption{Pan-European storage (top) and pipeline (bottom) capacity expansion results for the considered range of uncertain storage cost parameters ($GR=1.25$), own illustration based on optimisation results.}
    \label{fig:sum_of_sto_pip_2050}
\end{figure}

\paragraph{{Capacity investments}}
The total investment decisions in H\textsubscript{2} storage and pipeline capacity for all ten cost scenarios can be observed in \Cref{fig:sum_of_sto_pip_2050}.
Except for United Kingdom, the storage repurposing potentials are fully exploited in the scenarios with additional market uptake restrictions. 
For pipelines, the capacity generally increases slightly, as indicated by the green bars. 
The share of repurposing capacity for storage ranges from 20.2\,\% in the low storage cost scenario to 28.8\,\% in the high storage cost scenario. 
The share of repurposed pipeline assets varies from 82.9\,\% to 77.5\,\%. 
The installed H\textsubscript{2} storage capacity ranges from 220.6\,TWh\textsubscript{th} to 195.2\,TWh\textsubscript{th}, and the transport capacity via pipelines ranges from 12.5\,TWh\textsubscript{th}/d to 15.4\,TWh\textsubscript{th}/d. 

\begin{figure*}[h]
\centering
  \includegraphics[width=0.98\linewidth]{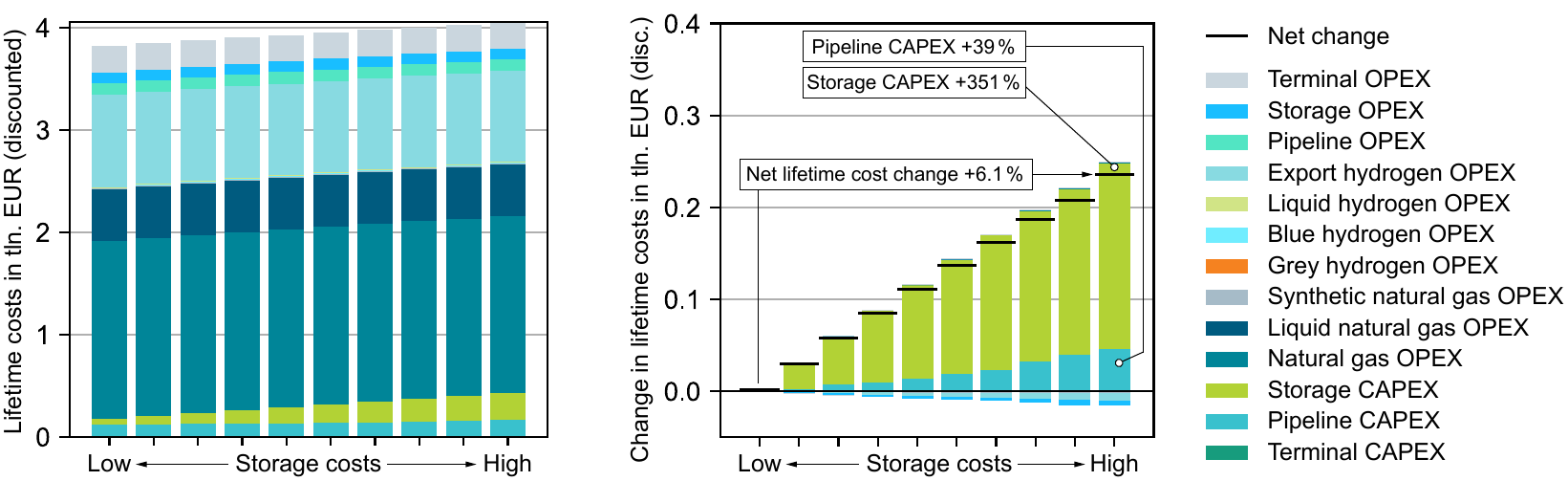}
  \caption{Comparison of resulting total and change in lifetime discounted costs for the considered low to high H\textsubscript{2} storage cost scenarios and the market growth rate variant $GR = 1.25$, own illustration based on optimisation results.}
   \label{fig:liftime_costs}
\end{figure*}

\paragraph{{Lifetime costs analysis}}
Because the objective function of the optimisation minimises the total lifetime costs, the resulting investment and operation costs are essential results.
The investment costs capture those for capacity expansion or repurposing decisions, and decommissioning costs (including lifetime compensation payments) for decommissioned infrastructure.  
The costs can also be separated into period-specific and lifetime investment costs, which are the sum over the entire multi-period planning horizon until the final state of the system is reached.   
Varying only one parameter in the optimisation, e.g. the storage investment costs, largely affects the solution and composition of the H\textsubscript{2} and CH\textsubscript{4} infrastructure, see \Cref{fig:liftime_costs}. 
In the high-cost scenario, the overall system costs increase by around 6.2\,\% compared to the low storage cost scenario.
\par
As more H\textsubscript{2} pipeline infrastructure is needed earlier in the transformation pathways, there is a extended need for LNG in the markets, see \Cref{fig:liftime_costs}. 
The largest share of the lifetime costs is incurred by natural gas imports and the H\textsubscript{2} produced outside of Europe, see the left plot in \Cref{fig:liftime_costs}. 

\begin{figure*}[h]
    \centering
    \includegraphics[width=0.98\linewidth]{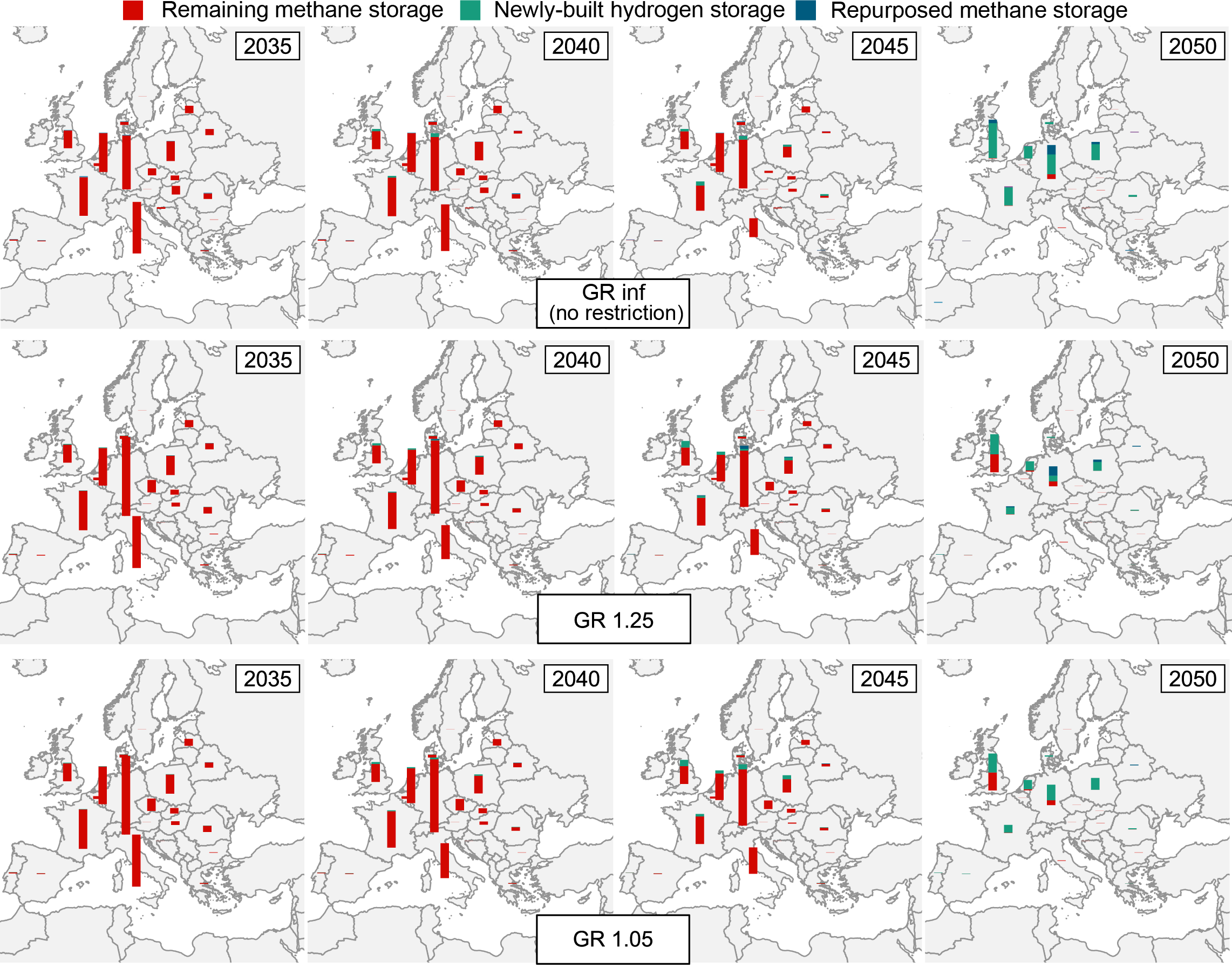}
    \caption{Impacts of different market growth rates (infinity, 1.25, 1.05) on installed H\textsubscript{2} storage capacity in Europe from 2035 to 2050 for the low-cost storage cost scenario, own illustration based on optimisation results.}
    \label{fig:gr_differences}
\end{figure*}

\subsection{Impacts of market growth restrictions}
\label{subsec:MarketGrowthUncertainty}
The model results demonstrate that the market growth rate plays a crucial role in determining the expansion of H\textsubscript{2} storage, as well as the repurposing and decommissioning of CH\textsubscript{4} storage facilities, as shown in \Cref{fig:gr_differences}.
Due to the continued consumption of CH\textsubscript{4} in the later periods, a strong shift in the trajectory is not feasible given stricter market growth rates.
The findings indicate that the market growth rate has a substantial influence on optimising infrastructure investments and prevents abrupt build-outs.
\par
\begin{figure*}
    \includegraphics[width=.5\linewidth]{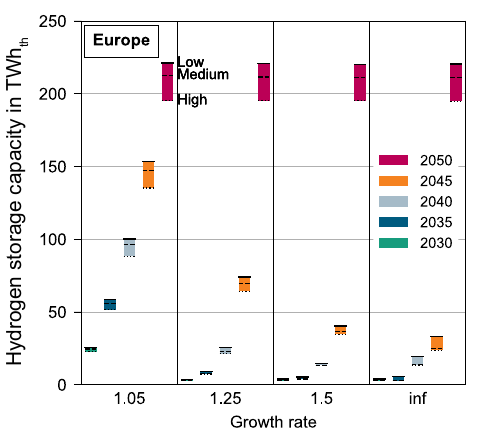}\hfill
    \includegraphics[width=.5\linewidth]{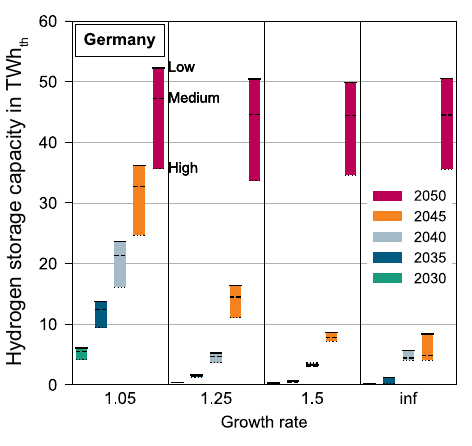}\hfill
    \caption{Total hydrogen (H\textsubscript{2}) storage demand results (including expansion, refurbishing, and decommissioning) for four market growth rate variations (1.05, 1.25, 1.5, and infinity) and three storage cost scenarios (Low, Medium, High) in Europe (left) and Germany (right), own illustration based on optimisation results.}
    \label{fig:Box_plot}
\end{figure*}

\Cref{fig:Box_plot} illustrates the coactive effects of varying market growth rate restrictions and storage investment costs for on H\textsubscript{2} storage demands in Europe and Germany. Note that the individual results for capacity expansion and refurbishing are given in \Cref{fig:Box_plot_neu_rep_europe,fig:Box_plot_neu_rep_germany}, respectively.
Comparing the results between Europe (left) and Germany (right) in \Cref{fig:Box_plot}, it is evident that the storage demand in Germany is strongly affected by the storage cost sensitivities, while the European total H\textsubscript{2} storage demands show smaller variations.
Between 2030 and 2045, the assumed growth rate determines how early the new and refurbishing investments need to pick up in order to meet the long-term H\textsubscript{2} demands.
Without market growth rate restrictions, H\textsubscript{2} storage demands will begin to emerge by 2035 or 2040.
However, if market growth restrictions are imposed, the modelling results exhibit storage build-outs between 2025 and 2030.
The growth rate restrictions mostly affect the path-dependent investment trajectories, whereas the storage costs have the greatest impact on the final H\textsubscript{2} storage demand by 2050.
\par
The gap between 2045 and 2050 is large for all variants except for the 1.05 growth rate variants. 
Even in the 1.25 variants, the installed capacity for hydrogen storage in Europe tripled in these five years.
The difference between 2045 and 2050 is even higher in the low-cost scenarios. 
To assume realistic market growth, a growth rate lower than 1.25 should be considered. 
\par
\Cref{fig:Box_plot_pip_europe} in the appendix shows Europe's resulting aggregated H\textsubscript{2} pipeline capacities, comparing the pipeline investment decisions for different H\textsubscript{2} storage costs and market growth rates.  
It is clear that the impact of the growth rates is much lower for pipelines than for storage, which may be due to the need for more exchange capacity within Europe in earlier years, as well as to more repurposing and fewer construction projects for new pipelines.

\subsection{Discussion}
\subsubsection{Summary and reflection}
The case study results demonstrate the possibility of integrating H\textsubscript{2} as a key piece into a climate-neutral energy system. 
As other studies focus on the final configuration \cite{v.MikuliczRadecki.2023,EUROPEANCOMMISSION.2022,Neumann.2023,Wietschel.2021}, the IMAGINE modelling framework also incorporates the path-dependent decisions involved in the CH\textsubscript{4} and H\textsubscript{2} gas infrastructure development over multiple planning periods.
Besides showing the interactions between different components of the gas infrastructure transformation, the approach allows us to shed light on two important uncertainties, i.e. H\textsubscript{2} storage costs and market uptake restrictions.
The findings reveal significant influences on the demand trajectories for hydrogen storage.
While the costs associated with hydrogen storage predominantly determine the final gas infrastructure configuration of a net-neutral energy system, restrictions on market uptake critically affect the intermediate hydrogen storage demands across Europe.
Additionally, the repurposing of natural gas infrastructure is crucial in all examined scenarios and nearly every country.
\par
The soft-linking approach of the SCOPE SD and IMAGINE models results in a more inefficient system than integrating all decisions in one model because investment and operation decisions in electricity capacities and infrastructure are made without consideration of the gas infrastructure and market.
However, due to varying technological readiness and policy environments, it is unknown whether full integration of all markets, including power, gas, and others, will eventually occur or remain elusive.
Hence, our soft-linking approach might represent a pragmatic alignment with real-world market and infrastructure development processes.
\par
Based on the scenario setup, we show that H\textsubscript{2} storage investments start as early as 2030 but become essential between 2045 and 2050, confirming the results in \cite{Frischmuth.2022b}.
H\textsubscript{2} pipeline investments start in 2025 and ramp up until 2050. 
These results align with the work presented in \cite{Victoria.2022}, where the H\textsubscript{2} networks appear after 2035. 
Because of the extreme H\textsubscript{2} storage investments in 2045 and 2050, the model was extended to investigate the uncertainty associated with market uptake restrictions. 
As evidenced by Great Britain's case, the market growth rate plays a critical role in repurposing and decommissioning gas infrastructure components.
The assumptions for the market growth rate exhibit substantial impacts, highlighting its relevance for multi-period expansion planning and the need for a better understanding and quantification of the market uptake processes and their restrictions.
A market growth rate restriction, which in our case study means lower than 1.25, should be used to represent technology deployment and transformation pathways more realistically in the models.
In addition to the uptake of the H\textsubscript{2} economy, individual storage and pipeline projects take time to build.
For example, in \cite{NRW.ENERGY.4CLIMATE.2022}, it is assumed that the construction of a new underground H\textsubscript{2} storage facility takes 15 years, and the repurposing takes up to five years. 
\par
Due to the key role of the MENA region, as well as the transit countries, the responsibility of building Europe's future H\textsubscript{2} network including environmental, social, and financial concerns need to be addressed in the pursuit of a just transition, e.g., see  \cite{LONERGAN2023113791}.
\par
In the study conducted in \cite{Neumann.2023}, European climate neutrality scenarios were examined, considering electricity and H\textsubscript{2} network trade-offs along with integrated electricity and H\textsubscript{2} markets. 
The authors find cost-optimal H\textsubscript{2} storage demands in Europe of up to 43\,TWh\textsubscript{th}.
By contrast, in \cite{Alibas.2024}, the demand for H\textsubscript{2} storage in Europe ranges between 143 and 268\,TWh\textsubscript{th}.
In \cite{Caglayan.2020}, the demand for H\textsubscript{2} storage is about 130\,TWh\textsubscript{th}.
Our scenarios involve H\textsubscript{2} storage investments ranging from 195 to 220\,TWh\textsubscript{th} until 2050.
There are a few important aspects to keep in mind when comparing the results with \cite{Neumann.2023}, which does not consider (green) H\textsubscript{2} imports from outside of Europe, implying that all H\textsubscript{2} (or derived fuels) will be produced within the European energy system.
While the overall H\textsubscript{2} consumption in the final planning period is at similar levels for all studies, the results obtained by our case study assume domestic electrolyser deployments of up to 410\,GW\textsubscript{el} as compared to 1250\,GW\textsubscript{el} in \cite{Neumann.2023}.
\par
Moreover, when considering the possibility of importing renewable (storable) fuels into Europe, several factors influence gas storage demands in Europe.
Weather-dependent export patterns increase the demand for gas storage, which can be balanced close to the production site, during gas transport (e.g. a shipping vessel), or by the importing region, close to the entry point, refinery, or consumption site.
Ultimately, the coordination of the electricity system with the gas infrastructure strongly drive the absolute H\textsubscript{2} storage demands.
However, for a given energy system development pathway, this work highlights the impacts of two distinct uncertainties --- i.e. H\textsubscript{2} storage costs and market growth restrictions of building new, refurbishing or decommissioning existing capacities --- on the path-dependent H\textsubscript{2} storage demands in the gas infrastructure transition.
\par
Another distinguishing factor is that our model and study incorporate refurbishing projects into the decision-making model.
The possibility of repurposing natural gas storage facilities at lower costs contributes to the higher H\textsubscript{2} storage requirements in our results. 
In addition, the share of repurposed natural gas pipelines ranges from 77\,\% to 83\,\% in 2050, while \cite{Neumann.2023,Jens.2021} find slightly lower shares around 69\,\%. 
One explanation for the higher share in our study is that we focus on the transport between market areas and do not consider the transmission and distribution pipelines within the individual market areas.
Furthermore, we consider imports from outside of Europe which use existing or repurposed import pipeline infrastructure and require fewer new expansion projects.
Regarding repurposed natural gas storage, our study shows a corresponding range of 20\,\% to 29\,\% in 2050.
\par
Another aspect relevant to our case study is the withdrawal and injection rates, which are critical factors in assessing H\textsubscript{2} storage development.
However, the reviewed literature provides little information on these parameters.
It is important to note that due to geological circumstances, these parameters are highly individual for every cavern storage.
A generalised injection and withdrawal to storage volume rate of 15.4\,GWh\textsubscript{th}/GW\textsubscript{th} is assumed, which is lower than that of the existing CH\textsubscript{4} cavern storage, see \cite{GIE.2021}. 
According to \cite{JuezLarre.2023}, storage injection and withdrawal for H\textsubscript{2} storage is 3-4 times faster than for CH\textsubscript{4} cavern storage. 

\subsubsection{Limitations}
Model-based analyses exhibit limitations due to methodological choices and specific assumptions.
The main aspects limiting the outcomes of this study are discussed below.
\par
Our IMAGINE modelling framework focuses on the overall system interactions between different components of the gas infrastructure transition, thus providing insights that are both accessible and relevant to a wide range of stakeholders, including policymakers, industry participants, and academics.
Ensuring that the multi-period planning problem remains both computationally tractable and comprehensible requires simplifications:
\begin{enumerate}
    \item we adopt a linear gas transport model and omit nonlinear and nonconvex characteristics of gas transport, including linepacking, which allows the system to store additional amounts of gas within the pipeline system itself;
    \item our market-based approach currently does not consider individual pipelines and internal congestions within the bidding zones;
    \item our model features a daily instead of a finer temporal resolution.
\end{enumerate}
These aspects can all impact the results of the obtained CH\textsubscript{4} and H\textsubscript{2} storage demands. Additional linepacking storage could reduce the demand for H\textsubscript{2} storage. In contrast, more detailed regional modelling of pipelines and demands can be expected to increase the demand for gas storage due to additional network congestion. However, both aspects will likely counteract, and the daily temporal resolution might attenuate some effects.
Therefore, despite the methodological choices, our model offers a reasonable compromise for exploring the strategic development of CH\textsubscript{4} and H\textsubscript{2} infrastructure across Europe and the potential import options. It balances detail with the broader system perspective, essential for informed decision-making to reflect the critical dynamics of Europe’s transformation of the gas infrastructure.
\par
Although there are clear advantages to considering gas network expansion and refurbishment on an individual asset basis, as in electricity network planning, e.g.~\cite{Kristiansen.2016, Vrana.2023}, several factors render this detailed approach currently difficult.
While spatial data on pipelines and gas storage, including their refurbishment potentials, is increasingly available, it is often incomplete or lacks the granularity required for such detailed modelling.
Additionally, the computational tractability of a detailed multi-period planning framework remains a significant challenge.
The complexity introduced by incorporating every individual pipeline and storage facility, including integer decision variables, into the model can lead to excessive computational demands, making it impractical for large-scale network analysis.
Therefore, despite the potential benefits, we adopt a more aggregated modelling approach to ensure manageability and computational efficiency while providing indicative results for the European gas infrastructure transformation.
\par
Moreover, from a system development perspective, the planning framework does not incorporate a strategic reserve for H\textsubscript{2} supply. 
Hence, the modelling framework considers the security of supply only in the sense that all given demands are met.
That said, extraordinary circumstances and events harming the system's resilience are not part of the analysis.
Therefore, the results indicate a lower bound from the supply security perspective.
\par
The model does consider the possibility of blending CH\textsubscript{4} and H\textsubscript{2} or partially repurposing the pipeline through compressor upgrades or pipeline reinforcements.
For model extensions, it may be relevant to consider the findings of \cite{Klatzer.2022}, \cite{KLATZER2024122264}, and \cite{Morales-Espana.2023}, who show relevant impacts on infrastructure expansion decisions.
\par
Due to high costs, the model does not consider the possibility of storing H\textsubscript{2} as a derivative, e.g., methanol. 
In \cite{Brown.2023}, it is shown that when storing H\textsubscript{2} in underground caverns is not feasible in some regions, storing H\textsubscript{2} derivatives becomes a viable option. It should be considered in future analyses as it could reduce the need for H\textsubscript{2} cavern storage in central Europe.
\par
Besides H\textsubscript{2} storage costs and market uptake restrictions, there are many other sources of uncertainty involved in developing an integrated energy system. 
The demand for H\textsubscript{2} in the power, industry, transport, and agriculture sectors is one crucial aspect that has not been the subject of this study.

\section{Conclusion}
\label{sec:summaryAndConclusion}
The case study is an integrated analysis of Europe's CH\textsubscript{4} and H\textsubscript{2} infrastructure transformation pathways by linking the pan-European energy system planning model SCOPE SD with the multi-period European gas market model IMAGINE.
Our sensitivity analysis exposes significant impacts on the multi-period transformation pathways due to storage cost and market growth restriction uncertainties associated with H\textsubscript{2} storage development.
The results exhibit strong impacts on H\textsubscript{2} storage demand trajectories.
While hydrogen storage costs predominantly affect the final state of the net-neutral energy system, the market uptake restrictions show critical impacts on the intermediate demands for European hydrogen storage.
Repurposing natural gas infrastructure remains essential in all scenarios and almost all countries.
\par
The insights have implications for several stakeholders involved in the gas infrastructure development.
Comprehensive scenario planning is required to explore a wide range of future possibilities.
Given the relevance of the investigated uncertainties and potential others, developing risk management strategies for system planners and operators is essential to facilitate a more resilient system transition.
Research and development efforts could prioritise reducing the risk by better understanding the uncertainties involved, e.g. hydrogen storage investment and refurbishing costs, injection and withdrawal rates, and future end-use demands for hydrogen.
The study's findings underscore the necessity of moderating market growth rates to support the H\textsubscript{2} economy's development effectively.
In order to achieve the required capacity by the target years of 2045 and 2050, it is crucial to commence the construction of H\textsubscript{2} storage and pipeline infrastructure imminently.
It is recommended that policymakers undertake comprehensive investigations into storage demands and allocation strategies.
This would facilitate the planning of early-stage storage projects across Europe, including initiatives such as the ``Wasserstoff-Kernnetz'' \cite{BenetzA.2024} pipeline in Germany.
\par
Given the expected low capacity utilisation in the early stages of these projects, it is essential to ensure long-term investment stability.
While they are of significant public interest and align with broader environmental and economic objectives, a policy approach focusing on strategic planning and targeted support could help facilitate this stability.
\par
Immediate avenues for future research include improved planning frameworks that treat uncertainties in an endogenous manner and make decisions under uncertainty to provide more robust decision support.
Further model improvements include higher spatial and temporal resolutions and incorporating more detailed gas infrastructure representations.

\section*{Acknowledgment}
The work of this paper has been performed as part of the project \href{https://www.iee.fraunhofer.de/de/projekte/suche/2021/dev-kopsys-2.html}{\textit{DeV-KopSys-2}}, which receives funding from the German Federal Ministry for Economic Affairs and Climate Action (BMWK) under funding reference numbers 16EM5008-1. Moreover, we thank Norman Gerhardt and Philipp Hahn for valuable comments during the drafting phase.

\bibliographystyle{unsrtnat}
\bibliography{refs}  

 
\appendix

\section{Supplementary information}
\begin{figure*}
    \includegraphics[width=.5\linewidth]{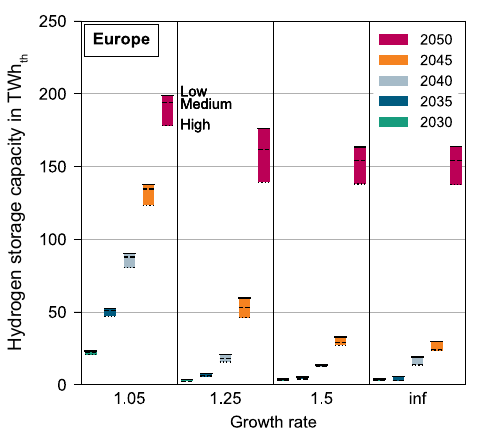}\hfill
    \includegraphics[width=.495\linewidth]{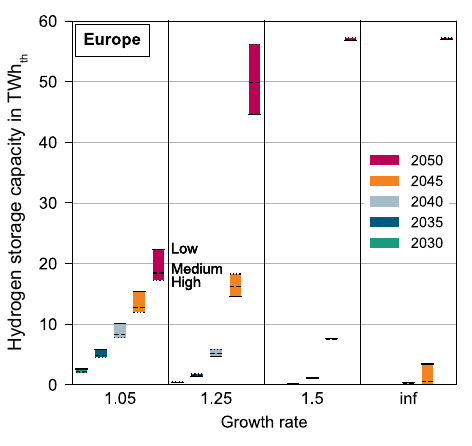}
    \caption{Newly-built (left) and repurposed (right) hydrogen (H\textsubscript{2}) storage decision results for four market growth rate variations (1.05, 1.25, 1.5, and infinity) and three storage cost scenarios (Low, Medium, High) in Europe, own illustration based on optimisation results.}
    \label{fig:Box_plot_neu_rep_europe}
\end{figure*}

\begin{figure*}
    \includegraphics[width=.5\linewidth]{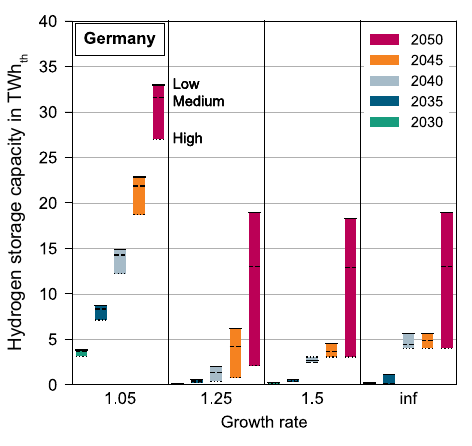}\hfill
    \includegraphics[width=.5\linewidth]{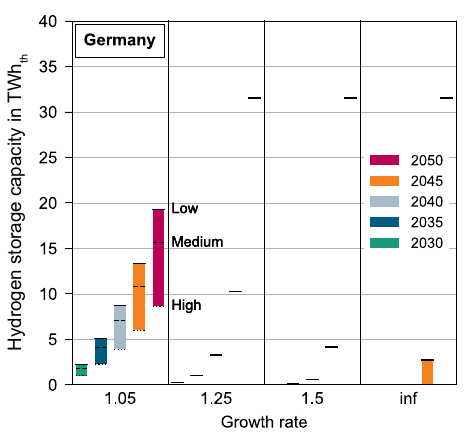}
    \caption{Newly-built (left) and repurposed (right) hydrogen (H\textsubscript{2}) storage decision results for four market growth rate variations (1.05, 1.25, 1.5, and infinity) and three storage cost scenarios (Low, Medium, High) in Germany, own illustration based on optimisation results.}
    \label{fig:Box_plot_neu_rep_germany}
\end{figure*}

The appendix contains supplementary material providing additional information on the IMAGINE framework and the case study's assumption and results.
\Cref{tab:PipelineCosts} gives an overview on the assumed costs for pipelines, storage and terminal.
\Cref{tab:HydrogenGenCom} and \Cref{tab:MethaneGenCom} shows results of the SCOPE SD model, which serve as input data for the IMAGINE model.
\Cref{tab:NaturalGasProCost} \Cref{tab:GreenHydrogenExport} provide assumptions on potentials and costs for CH\textsubscript{4} and H\textsubscript{2} sources in the model. 
\Cref{fig:Box_plot_neu_rep_europe} and \Cref{fig:Box_plot_neu_rep_germany} offer additional insight into the impact of market growth rates and storage costs for newly-built and repurposed H\textsubscript{2} storage in Europe and Germany. 


\begin{figure}[h]
\centering
    \includegraphics[width=0.5\linewidth]{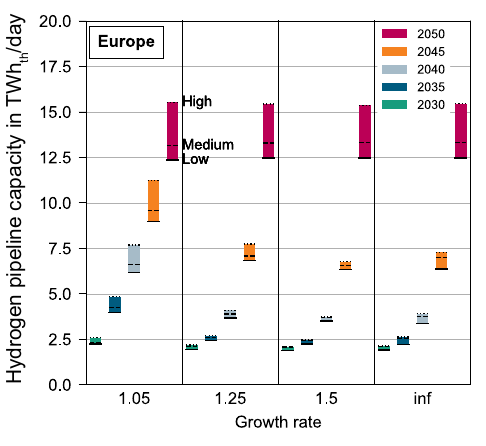}\hfill
    \caption{Hydrogen (H\textsubscript{2}) pipeline decision results for four market growth rate variations (1.05, 1.25, 1.5, and infinity) and three storage cost scenarios (Low, Medium, High) in Europe, own illustration based on optimisation results.}
    \label{fig:Box_plot_pip_europe}
\end{figure}



\begin{table*}[h]
    \centering
    \caption{Pipeline, terminal and storage investment and fixed operation costs, own calculation based on \cite{Backbone.2020,BDI.2022,DEA.2020,FlorianSchreinerMatiaRiemerJakobWachsmuth.}}
    \begin{tabular}{l r r r }
        \toprule
            & \multicolumn{1}{l}{Investment cost} & \multicolumn{1}{l}{Fixed operation cost}  \\
            Technology  & \multicolumn{1}{l}{in EUR/km/$\frac{\text{GWh}_\text{th}}{\text{d}}$} & \multicolumn{1}{l}{in EUR/km/$\frac{\text{GWh}_\text{th}}{\text{d}}$/yr}     \\
        \midrule
        Pipeline    CH\textsubscript{4} new & 6,828  & 85       \\
        Pipeline    H\textsubscript{2} new & 10,924  &  137       \\
        Pipeline    H\textsubscript{2} retrofit & 3,690 & 46       \\
        \midrule
        & \multicolumn{1}{l}{in EUR/GWh\textsubscript{th}} & \multicolumn{1}{l}{in EUR/GWh\textsubscript{th}/yr}     \\
        \midrule
        Storage    H\textsubscript{2} new (low-cost) & 269,393 & 5,388       \\
        Storage    H\textsubscript{2} new (+50)\% & 404,088 & 8,082       \\
        Storage    H\textsubscript{2} new (+100)\%  & 538,783 & 10,776       \\
        Storage    H\textsubscript{2} new (...)\%  & ... & ...       \\
        Storage    H\textsubscript{2} new (+350)\% & 1,212,263 & 24,245       \\
        Storage    H\textsubscript{2} new (+400)\% & 1,346,959 & 26,939      \\
        Storage    H\textsubscript{2} new (high-cost) & 1,500,000 & 30,000       \\
        Storage    H\textsubscript{2} retrofit (low-cost) & 45,472 & 909       \\
        Storage    H\textsubscript{2} retrofit (...)\%  & ... & ...       \\
        Storage    H\textsubscript{2} retrofit (high-cost) & 253,194 & 5,063       \\
        \midrule
        & \multicolumn{1}{l}{in EUR/GWh\textsubscript{th}/d} & \multicolumn{1}{l}{in EUR/GWh\textsubscript{th}/d/yr}     \\
        \midrule
        Terminal    CH\textsubscript{4} new & 5,040,000 & 100,800      \\
        Terminal    H\textsubscript{2} new & 2,646,000 & 52,920       \\
        Terminal    H\textsubscript{2} retrofit & 882,000 & 52,920       \\
        \bottomrule
    \end{tabular}
    \label{tab:PipelineCosts}
\end{table*}

\begin{table*}[h]
\footnotesize
    \centering
    \caption{Assumptions and results of the SCOPE SD model for H\textsubscript{2} generation an consumption within Europe.} 
\begin{tabular}{lrrrrrrr}
\toprule
& \multicolumn{7}{c}{Domestic H\textsubscript{2} consumption in TWh\textsubscript{th}/yr / Domestic green H\textsubscript{2} production in TWh\textsubscript{th}/yr} \\
Country & 2020 & 2025 & 2030 & 2035 & 2040 & 2045 & 2050\\
\hline
Austria           & 0.0 / 0.0 & 1.2 / 0.0 & 7.5 / 1.1 & 11.7 / 2.5 & 22.7 / 2.3 & 25.9 / 3.0 & 36.7 / 3.4\\
Belgium           & 0.0 / 0.0 & 1.7 / 0.0 & 13.3 / 3.3 & 24.8 / 5.0 & 59.6 / 4.6 & 87.3 / 5.1 & 138.2 / 6.1\\
Bulgaria          & 0.0 / 0.0 & 0.7 / 0.0 & 3.3 / 0.7 & 3.8 / 4.2 & 8.4 / 13.4 & 7.6 / 16.8 & 9.9 / 12.4\\
Croatia           & 0.0 / 0.0 & 0.3 / 0.0 & 1.9 / 0.3 & 1.8 / 0.6 & 1.7 / 0.7 & 2.7 / 5.6 & 3.6 / 6.9\\
Cyprus    & 0.0 / 0.0 & 0.0 / 0.0 & 0.0 / 0.0 & 0.0 / 0.0 & 0.1 / 0.0 & 0.2 / 0.0 & 0.2 / 0.0\\
Czech Republic    & 0.0 / 0.0 & 1.2 / 0.0 & 7.9 / 0.4 & 13.3 / 2.2 & 17.8 / 2.2 & 33.4 / 5.9 & 53.5 / 20.0\\
Denmark           & 0.0 / 0.0 & 0.6 / 0.0 & 2.5 / 7.7 & 3.0 / 12.1 & 2.9 / 14.0 & 5.8 / 18.0 & 7.8 / 11.6\\
Estonia           & 0.0 / 0.0 & 0.1 / 0.0 & 0.5 / 0.1 & 0.6 / 0.3 & 0.5 / 2.9 & 1.2 / 3.8 & 1.8 / 2.8\\
Finland           & 0.0 / 0.0 & 0.4 / 0.0 & 6.6 / 1.5 & 7.5 / 2.8 & 10.3 / 6.8 & 17.5 / 7.5 & 26.6 / 10.3\\
France    & 0.0 / 0.0 & 0.9 / 0.0 & 21.0 / 15.3 & 38.1 / 32.6 & 87.9 / 78.9 & 146.4 / 156.6 & 211.6 / 190.1\\
Germany           & 0.0 / 0.0 & 20.8 / 0.0 & 65.0 / 24.0 & 100.3 / 70.1 & 175.6 / 175.7 & 317.0 / 192.2 & 438.9 / 225.8\\
Greece    & 0.0 / 0.0 & 0.5 / 0.0 & 5.0 / 0.7 & 5.5 / 12.6 & 6.2 / 20.2 & 9.6 / 28.0 & 12.5 / 15.4\\
Hungary           & 0.0 / 0.0 & 0.9 / 0.0 & 4.9 / 0.8 & 8.1 / 2.0 & 14.5 / 10.4 & 24.1 / 22.0 & 33.7 / 24.8\\
Ireland           & 0.0 / 0.0 & 0.3 / 0.0 & 1.4 / 0.6 & 1.6 / 9.1 & 2.2 / 11.1 & 3.7 / 15.2 & 6.5 / 9.6\\
Italy     & 0.0 / 0.0 & 5.3 / 0.0 & 27.4 / 4.2 & 36.2 / 4.8 & 69.3 / 10.3 & 92.4 / 53.5 & 139.3 / 102.5\\
Latvia    & 0.0 / 0.0 & 0.2 / 0.0 & 0.7 / 0.0 & 0.9 / 1.2 & 0.6 / 3.3 & 1.2 / 4.3 & 1.7 / 2.7\\
Luxembourg        & 0.0 / 0.0 & 0.1 / 0.0 & 0.4 / 0.0 & 0.6 / 0.0 & 2.3 / 0.0 & 2.3 / 0.0 & 3.9 / 0.0\\
Lithuania         & 0.0 / 0.0 & 0.1 / 0.0 & 1.7 / 0.1 & 2.1 / 3.7 & 2.9 / 4.9 & 6.7 / 6.5 & 9.5 / 4.1\\
Netherlands       & 0.0 / 0.0 & 1.9 / 0.0 & 20.1 / 33.1 & 31.7 / 33.1 & 61.4 / 33.1 & 109.0 / 46.3 & 178.2 / 76.7\\
Norway    & 0.0 / 0.0 & 0.3 / 0.0 & 1.5 / 0.0 & 2.5 / 0.0 & 4.7 / 0.0 & 10.7 / 0.0 & 15.5 / 0.0\\
Poland    & 0.0 / 0.0 & 0.2 / 0.0 & 7.8 / 0.4 & 13.2 / 2.7 & 36.6 / 5.2 & 60.1 / 20.0 & 144.4 / 57.1\\
Portugal          & 0.0 / 0.0 & 0.9 / 0.0 & 5.4 / 9.2 & 6.4 / 9.2 & 10.8 / 12.5 & 13.8 / 16.6 & 17.8 / 18.9\\
Romania           & 0.0 / 0.0 & 1.2 / 0.0 & 6.5 / 1.7 & 8.8 / 12.2 & 27.2 / 33.2 & 21.8 / 48.3 & 30.9 / 31.1\\
Slovakia          & 0.0 / 0.0 & 0.7 / 0.0 & 5.8 / 0.2 & 9.0 / 1.0 & 15.7 / 1.2 & 19.3 / 5.8 & 26.1 / 14.6\\
Slovenia          & 0.0 / 0.0 & 0.4 / 0.0 & 1.4 / 0.3 & 1.9 / 0.7 & 1.9 / 0.8 & 3.2 / 1.0 & 4.2 / 1.2\\
Spain     & 0.0 / 0.0 & 0.3 / 0.0 & 20.3 / 21.6 & 23.2 / 18.0 & 52.4 / 69.0 & 65.4 / 85.9 & 87.1 / 96.2\\
Sweden    & 0.0 / 0.0 & 0.5 / 0.0 & 5.6 / 8.5 & 8.2 / 9.5 & 12.7 / 10.2 & 24.5 / 41.6 & 34.8 / 51.7\\
Switzerland       & 0.0 / 0.0 & 1.2 / 0.0 & 7.1 / 0.0 & 11.4 / 0.0 & 21.2 / 0.0 & 25.1 / 0.0 & 36.7 / 0.0\\
United Kingdom    & 0.0 / 0.0 & 5.9 / 0.0 & 28.4 / 6.8 & 33.8 / 10.1 & 52.4 / 61.7 & 92.8 / 60.8 & 185.7 / 192.7\\
\midrule
EU-28 & 0.0 / 0.0 & 48.8 / 0.0 & 280.9 / 142.6 & 410.0 / 262.3 & 782.5 / 588.6 & 1230.7 / 870.3 & 1897.3 / 1188.7\\
\bottomrule
\end{tabular}
\label{tab:HydrogenGenCom}
\end{table*}

\begin{table*}[h]
\footnotesize
    \centering
    \caption{Assumptions and results of the SCOPE SD model for CH\textsubscript{4} consumption and fixed bio methane generation within Europe.} 
\begin{tabular}{lrrrrrrr}
\toprule
& \multicolumn{7}{c}{Domestic CH\textsubscript{4} consumption in TWh\textsubscript{th}/yr / Domestic Bio CH\textsubscript{4} production in TWh\textsubscript{th}/yr} \\
Country & 2020 & 2025 & 2030 & 2035 & 2040 & 2045 & 2050\\
\hline
Albania           & 0.4 / 0.0 & 0.3 / 0.0 & 0.1 / 0.0 & 0.0 / 0.0 & 0.0 / 0.0 & 0.0 / 0.0 & 0.0 / 0.0\\
Austria           & 83.7 / 0.3 & 77.8 / 0.5 & 57.8 / 0.8 & 49.1 / 1.1 & 28.3 / 1.4 & 10.7 / 1.7 & 0.0 / 0.0\\
Belgium           & 114.7 / 0.0 & 126.5 / 0.6 & 96.6 / 1.2 & 85.5 / 1.7 & 59.7 / 2.3 & 32.7 / 2.9 & 0.6 / 0.6\\
Bulgaria          & 20.5 / 0.0 & 20.0 / 0.2 & 24.4 / 0.4 & 11.4 / 0.6 & 3.3 / 0.8 & 0.8 / 1.0 & 0.1 / 0.1\\
Croatia           & 19.0 / 0.0 & 18.6 / 0.1 & 14.1 / 0.2 & 11.1 / 0.2 & 5.9 / 0.3 & 1.5 / 0.4 & 0.1 / 0.1\\
Czech Republic    & 73.2 / 0.0 & 72.6 / 0.5 & 69.9 / 1.1 & 53.6 / 1.6 & 40.5 / 2.2 & 22.9 / 2.7 & 0.0 / 0.0\\
Denmark           & 33.0 / 1.6 & 25.1 / 1.8 & 21.1 / 1.9 & 16.0 / 2.1 & 4.1 / 2.3 & 1.8 / 9.1 & 0.0 / 0.0\\
Estonia           & 10.0 / 0.1 & 8.3 / 0.1 & 6.9 / 0.1 & 4.8 / 0.2 & 2.7 / 0.2 & 0.9 / 0.2 & 0.0 / 0.0\\
Finland           & 88.0 / 0.2 & 77.7 / 0.6 & 66.9 / 1.1 & 46.4 / 1.6 & 18.4 / 2.0 & 5.4 / 2.5 & 0.0 / 0.0\\
France    & 370.7 / 2.3 & 355.7 / 5.2 & 267.6 / 8.1 & 210.2 / 11.1 & 153.5 / 14.0 & 53.8 / 16.9 & 2.8 / 2.8\\
Germany           & 754.2 / 11.0 & 704.1 / 12.1 & 514.7 / 13.1 & 375.1 / 14.2 & 263.7 / 15.2 & 132.2 / 16.3 & 35.8 / 35.8\\
Greece    & 31.7 / 0.0 & 47.5 / 0.9 & 50.1 / 1.8 & 15.5 / 2.6 & 5.5 / 3.5 & 0.9 / 15.3 & 0.2 / 0.2\\
Hungary           & 68.4 / 0.1 & 69.4 / 0.5 & 70.9 / 1.0 & 41.2 / 1.4 & 23.8 / 1.8 & 8.9 / 2.3 & 0.4 / 0.4\\
Ireland           & 43.6 / 0.0 & 32.8 / 0.5 & 18.9 / 0.9 & 12.1 / 1.3 & 6.0 / 1.7 & 3.3 / 6.4 & 0.2 / 0.2\\
Italy     & 464.2 / 1.6 & 468.1 / 3.2 & 292.5 / 4.8 & 247.1 / 6.5 & 118.0 / 8.1 & 55.6 / 9.7 & 1.8 / 1.8\\
Latvia    & 10.2 / 0.0 & 11.5 / 0.0 & 9.3 / 0.1 & 4.8 / 0.1 & 2.7 / 0.1 & 0.7 / 0.2 & 0.0 / 0.0\\
Luxembourg        & 5.6 / 0.1 & 6.3 / 0.1 & 4.3 / 0.1 & 4.2 / 0.1 & 3.3 / 0.2 & 1.7 / 0.2 & 0.0 / 0.0\\
Lithuania         & 24.7 / 0.0 & 25.0 / 0.2 & 23.0 / 0.4 & 18.0 / 0.6 & 8.3 / 0.7 & 2.0 / 4.4 & 0.0 / 0.0\\
Netherlands       & 238.6 / 2.2 & 240.5 / 2.5 & 183.0 / 2.9 & 150.0 / 3.3 & 105.7 / 3.7 & 47.9 / 4.0 & 1.0 / 1.0\\
Norway    & 32.8 / 0.2 & 31.6 / 0.4 & 28.5 / 0.5 & 19.1 / 0.7 & 2.8 / 0.9 & 0.8 / 1.1 & 0.0 / 0.0\\
Poland    & 218.3 / 0.0 & 193.6 / 1.9 & 282.0 / 3.8 & 241.8 / 5.7 & 179.9 / 7.6 & 125.9 / 9.6 & 0.1 / 0.1\\
Portugal          & 54.8 / 0.0 & 35.2 / 0.7 & 23.4 / 1.4 & 13.0 / 2.1 & 6.7 / 2.9 & 0.8 / 3.6 & 0.1 / 0.1\\
Romania           & 72.9 / 0.0 & 66.1 / 0.6 & 68.7 / 1.2 & 38.1 / 1.8 & 15.8 / 2.4 & 5.4 / 4.1 & 0.3 / 0.3\\
Serbia    & 25.1 / 0.0 & 16.7 / 0.0 & 8.4 / 0.0 & 0.0 / 0.0 & 0.0 / 0.0 & 0.0 / 0.0 & 0.0 / 0.0\\
Slovakia          & 45.9 / 0.0 & 41.9 / 0.2 & 40.7 / 0.4 & 26.6 / 0.7 & 17.6 / 0.9 & 7.7 / 1.1 & 0.2 / 0.2\\
Slovenia          & 10.7 / 0.0 & 10.8 / 0.0 & 8.1 / 0.1 & 7.1 / 0.1 & 4.0 / 0.2 & 1.5 / 0.2 & 0.1 / 0.1\\
Spain     & 333.6 / 0.1 & 164.4 / 2.6 & 95.9 / 5.2 & 58.7 / 7.7 & 30.5 / 10.2 & 4.8 / 12.8 & 0.8 / 0.8\\
Sweden    & 96.4 / 3.7 & 47.1 / 3.5 & 52.6 / 3.3 & 49.9 / 3.1 & 9.2 / 2.9 & 2.9 / 2.7 & 0.0 / 0.0\\
Switzerland       & 58.1 / 0.3 & 68.6 / 0.6 & 47.6 / 0.8 & 45.8 / 1.1 & 36.4 / 1.3 & 12.9 / 1.6 & 0.5 / 0.5\\
United Kingdom    & 618.1 / 4.2 & 422.8 / 8.5 & 317.2 / 12.8 & 209.2 / 17.1 & 157.3 / 21.3 & 115.6 / 25.6 & 3.4 / 3.4\\
\midrule
EU-28 & 4021.1 / 28.0 & 3486.6 / 48.6 & 2765.2 / 69.5 & 2065.4 / 90.4 & 1313.6 / 111.1 & 662.0 / 158.6 & 48.5 / 48.5\\
\bottomrule
\end{tabular}
    \label{tab:MethaneGenCom}
\end{table*}

\begin{table*}[h]
\footnotesize
    \centering
    \caption{Assumptions on natural gas production potentials and costs (without CO\textsubscript{2} costs), based on \cite{eurostat.2022,BMWK.2021,DIWBerlin.2019}.} 
\begin{tabular}{lrrrrrrr}
\toprule
& \multicolumn{7}{c}{Natural gas production potentials in TWh\textsubscript{th}/yr / Natural gas production costs in EUR/MWh\textsubscript{th}} \\
Country & 2020 & 2025 & 2030 & 2035 & 2040 & 2045 & 2050\\
\hline
Albania           & 0.4 / 18.8 & 0.3 / 18.8 & 0.1 / 18.8 & 0.0 / 18.8 & 0.0 / 18.8 & 0.0 / 18.8 & 0.0 / 18.8\\
Algeria           & 299.0 / 12.9 & 299.0 / 12.9 & 299.0 / 12.9 & 299.0 / 12.9 & 299.0 / 12.9 & 299.0 / 12.9 & 299.0 / 12.9\\
Austria           & 19.5 / 19.4 & 13.0 / 19.4 & 6.5 / 19.4 & 0.0 / 19.4 & 0.0 / 19.4 & 0.0 / 19.4 & 0.0 / 19.4\\
Belgium           & 21.5 / 19.4 & 14.4 / 19.4 & 7.2 / 19.4 & 0.0 / 19.4 & 0.0 / 19.4 & 0.0 / 19.4 & 0.0 / 19.4\\
Bulgaria          & 0.3 / 20.0 & 0.2 / 20.0 & 0.1 / 20.0 & 0.0 / 20.0 & 0.0 / 20.0 & 0.0 / 20.0 & 0.0 / 20.0\\
Croatia           & 11.8 / 20.0 & 7.9 / 20.0 & 3.9 / 20.0 & 0.0 / 20.0 & 0.0 / 20.0 & 0.0 / 20.0 & 0.0 / 20.0\\
Cyprus    & 0.0 / 11.8 & 0.0 / 11.8 & 0.0 / 11.8 & 0.0 / 11.8 & 0.0 / 11.8 & 0.0 / 11.8 & 0.0 / 11.8\\
Czech Republic    & 2.1 / 18.8 & 1.4 / 18.8 & 0.7 / 18.8 & 0.0 / 18.8 & 0.0 / 18.8 & 0.0 / 18.8 & 0.0 / 18.8\\
Denmark           & 43.1 / 20.0 & 28.7 / 20.0 & 14.4 / 20.0 & 0.0 / 20.0 & 0.0 / 20.0 & 0.0 / 20.0 & 0.0 / 20.0\\
Estonia           & 0.0 / 24.7 & 0.0 / 24.7 & 0.0 / 24.7 & 0.0 / 24.7 & 0.0 / 24.7 & 0.0 / 24.7 & 0.0 / 24.7\\
Finland           & 0.0 / 27.6 & 0.0 / 27.6 & 0.0 / 27.6 & 0.0 / 27.6 & 0.0 / 27.6 & 0.0 / 27.6 & 0.0 / 27.6\\
France    & 10.1 / 19.4 & 6.7 / 19.4 & 3.4 / 19.4 & 0.0 / 19.4 & 0.0 / 19.4 & 0.0 / 19.4 & 0.0 / 19.4\\
Germany           & 70.5 / 18.8 & 70.5 / 18.8 & 0.0 / 18.8 & 0.0 / 18.8 & 0.0 / 18.8 & 0.0 / 18.8 & 0.0 / 18.8\\
Greece    & 0.1 / 20.0 & 0.1 / 20.0 & 0.0 / 20.0 & 0.0 / 20.0 & 0.0 / 20.0 & 0.0 / 20.0 & 0.0 / 20.0\\
Hungary           & 17.1 / 18.8 & 11.4 / 18.8 & 5.7 / 18.8 & 0.0 / 18.8 & 0.0 / 18.8 & 0.0 / 18.8 & 0.0 / 18.8\\
Ireland           & 32.0 / 20.6 & 21.3 / 20.6 & 10.7 / 20.6 & 0.0 / 20.6 & 0.0 / 20.6 & 0.0 / 20.6 & 0.0 / 20.6\\
Italy     & 51.9 / 18.8 & 34.6 / 18.8 & 17.3 / 18.8 & 0.0 / 18.8 & 0.0 / 18.8 & 0.0 / 18.8 & 0.0 / 18.8\\
Latvia    & 0.0 / 25.9 & 0.0 / 25.9 & 0.0 / 25.9 & 0.0 / 25.9 & 0.0 / 25.9 & 0.0 / 25.9 & 0.0 / 25.9\\
Lithuania         & 0.0 / 25.9 & 0.0 / 25.9 & 0.0 / 25.9 & 0.0 / 25.9 & 0.0 / 25.9 & 0.0 / 25.9 & 0.0 / 25.9\\
Libya     & 62.7 / 12.9 & 62.7 / 12.9 & 62.7 / 12.9 & 62.7 / 12.9 & 62.7 / 12.9 & 62.7 / 12.9 & 62.7 / 12.9\\
Netherlands       & 488.5 / 18.2 & 488.5 / 18.2 & 488.5 / 18.2 & 488.5 / 18.2 & 488.5 / 18.2 & 0.0 / 18.2 & 0.0 / 18.2\\
Norway    & 1237.6 / 12.9 & 1237.6 / 12.9 & 1237.6 / 12.9 & 1237.6 / 12.9 & 0.0 / 12.9 & 0.0 / 12.9 & 0.0 / 12.9\\
Poland    & 40.3 / 18.8 & 26.9 / 18.8 & 13.4 / 18.8 & 0.0 / 18.8 & 0.0 / 18.8 & 0.0 / 18.8 & 0.0 / 18.8\\
Portugal          & 0.0 / 21.2 & 0.0 / 21.2 & 0.0 / 21.2 & 0.0 / 21.2 & 0.0 / 21.2 & 0.0 / 21.2 & 0.0 / 21.2\\
Romania           & 99.6 / 18.2 & 66.4 / 18.2 & 33.2 / 18.2 & 0.0 / 18.2 & 0.0 / 18.2 & 0.0 / 18.2 & 0.0 / 18.2\\
Russia    & 1285.9 / 9.4 & 0.0 / 28.2 & 0.0 / 28.2 & 0.0 / 28.2 & 0.0 / 28.2 & 0.0 / 28.2 & 0.0 / 28.2\\
Serbia    & 4.2 / 15.3 & 2.8 / 15.3 & 1.4 / 15.3 & 0.0 / 15.3 & 0.0 / 15.3 & 0.0 / 15.3 & 0.0 / 15.3\\
Slovakia          & 0.9 / 20.0 & 0.6 / 20.0 & 0.3 / 20.0 & 0.0 / 20.0 & 0.0 / 20.0 & 0.0 / 20.0 & 0.0 / 20.0\\
Slovenia          & 0.9 / 20.0 & 0.6 / 20.0 & 0.3 / 20.0 & 0.0 / 20.0 & 0.0 / 20.0 & 0.0 / 20.0 & 0.0 / 20.0\\
Spain     & 4.3 / 21.2 & 2.8 / 21.2 & 1.4 / 21.2 & 0.0 / 21.2 & 0.0 / 21.2 & 0.0 / 21.2 & 0.0 / 21.2\\
Sweden    & 0.0 / 22.4 & 0.0 / 22.4 & 0.0 / 22.4 & 0.0 / 22.4 & 0.0 / 22.4 & 0.0 / 22.4 & 0.0 / 22.4\\
Turkey    & 5.4 / 19.1 & 5.4 / 19.1 & 5.4 / 19.1 & 5.4 / 19.1 & 5.4 / 19.1 & 5.4 / 19.1 & 5.4 / 19.1\\
United Kingdom    & 406.8 / 18.8 & 406.8 / 18.8 & 0.0 / 18.8 & 0.0 / 18.8 & 0.0 / 18.8 & 0.0 / 18.8 & 0.0 / 18.8\\
\midrule
sum /mean & 4216.5 / 19.1 & 2810.6 / 19.7 & 2213.2 / 19.7 & 2093.2 / 19.7 & 855.6 / 19.7 & 367.1 / 19.7 & 367.1 / 19.7\\
\bottomrule
\end{tabular}
    \label{tab:NaturalGasProCost}
\end{table*}

\begin{table*}[h]
\footnotesize
    \centering
    \caption{Assumptions on green hydrogen (H\textsubscript{2}) export potentials and costs, based on Fraunhofer IEE's \textit{Power-to-X atlas}~\cite{FraunhoferIEE.2021}.} 
\begin{tabular}{lrrrrrrr}
\toprule
& \multicolumn{7}{c}{H\textsubscript{2} production (export) potential in TWh\textsubscript{th}/yr / H\textsubscript{2} production (export) costs in EUR/MWh\textsubscript{th}} \\
Country & 2020 & 2025 & 2030 & 2035 & 2040 & 2045 & 2050\\
\hline
Algeria           & 0.0 / 1045.0 & 0.0 / 964.8 & 157.0 / 104.5 & 157.0 / 96.5 & 157.0 / 88.5 & 157.0 / 80.4 & 157.0 / 72.4\\
Egypt     & 0.0 / 793.0 & 0.0 / 764.2 & 4180.0 / 79.3 & 4180.0 / 76.4 & 4180.0 / 73.5 & 4180.0 / 70.7 & 4180.0 / 67.8\\
Libya     & 0.0 / 1106.0 & 0.0 / 1003.0 & 3649.0 / 110.6 & 3649.0 / 100.3 & 3649.0 / 90.0 & 3649.0 / 79.7 & 3649.0 / 69.4\\
Morocco           & 0.0 / 762.0 & 0.0 / 729.8 & 376.0 / 76.2 & 376.0 / 73.0 & 376.0 / 69.8 & 376.0 / 66.5 & 376.0 / 63.3\\
Tunisia           & 0.0 / 846.0 & 0.0 / 817.5 & 265.0 / 84.6 & 265.0 / 81.8 & 265.0 / 78.9 & 265.0 / 76.0 & 265.0 / 73.2\\
Saudi Arabia      & 0.0 / 1064.0 & 0.0 / 975.2 & 872.0 / 106.4 & 872.0 / 97.5 & 872.0 / 88.7 & 872.0 / 79.8 & 872.0 / 70.9\\
\midrule
sum / mean & 0.0 / 936.0 & 0.0 / 875.8 & 9499.0 / 93.6 & 9499.0 / 87.6 & 9499.0 / 81.6 & 9499.0 / 75.5 & 9499.0 / 69.5\\
\bottomrule
\end{tabular}
    \label{tab:GreenHydrogenExport}
\end{table*}

\end{document}